\documentclass[twocolumn,english,reprint,superscriptaddress,longbibliography,pra]{revtex4-1}

\usepackage[T1]{fontenc}
\usepackage[latin9]{inputenc}
\usepackage{color}
\usepackage{bm}
\usepackage{amstext}
\usepackage{amssymb}
\usepackage{graphicx}
\usepackage{booktabs}
\usepackage{tabularx}
\usepackage{amsmath,mathrsfs}
\usepackage{enumitem}
\makeatletter
\usepackage{multibib}
\usepackage{listings}
\usepackage{mathtools}
\usepackage{braket}

\usepackage{babel}
\usepackage{bbold}
\usepackage{mathrsfs}
\usepackage{hyperref}
\hypersetup{
 linktocpage = true,
     colorlinks,
    citecolor=blue,
    filecolor=black,
    linkcolor=blue,		
    urlcolor=blue
}
\newcommand{\BeH}{BeH$_2$}
\newcommand{\tr}{\textrm{Tr}}

\usepackage{xcolor}
\definecolor{codegreen}{rgb}{0,0.6,0}
\definecolor{codegray}{rgb}{0.5,0.5,0.5}
\definecolor{codepurple}{rgb}{0.58,0,0.82}
\definecolor{backcolour}{rgb}{0.95,0.95,0.92}
\lstdefinestyle{mystyle}{
    backgroundcolor=\color{backcolour},   
    commentstyle=\color{codegreen},
    keywordstyle=\color{magenta},
    numberstyle=\tiny\color{codegray},
    stringstyle=\color{codepurple},
    basicstyle=\ttfamily\footnotesize,
    breakatwhitespace=false,         
    breaklines=true,                 
    captionpos=b,                    
    keepspaces=true,                 
    numbers=left,                    
    numbersep=5pt,                  
    showspaces=false,                
    showstringspaces=false,
    showtabs=false,                  
    tabsize=2
}
\usepackage{xpatch}

\makeatletter
\patchcmd{\@ssect@ltx}
    {\addcontentsline{toc}{#1}{\protect\numberline{}#8}}
    {}
    {}
    {}
\makeatother

\begin{document}

\title{Neural networks in quantum many-body physics: a hands-on tutorial}

\author{Juan Carrasquilla}
\affiliation{Vector Institute, MaRS Centre, Toronto, Ontario, M5G 1M1, Canada}
\author{Giacomo Torlai}
\email{gttorlai@amazon.com}
\thanks{\\This work has been done before Giacomo Torlai joined Amazon.}
\affiliation{AWS Center for Quantum Computing, Pasadena, CA 91125, USA}
\affiliation{Center for Computational Quantum Physics, Flatiron Institute, New York, NY 10010, USA}

\begin{abstract}
Over the past years, machine learning has emerged as a powerful computational tool to tackle complex problems over a broad range of scientific disciplines. In particular, artificial neural networks have been successfully deployed to mitigate the exponential complexity often encountered in quantum many-body physics, the study of properties of quantum systems built out of a large number of interacting particles. In this Article, we overview some applications of machine learning in condensed matter physics and quantum information, with particular emphasis on hands-on tutorials serving as a quick-start for a newcomer to the field. We present supervised machine learning with convolutional neural networks to learn a phase transition, unsupervised learning with restricted Boltzmann machines to perform quantum tomography, and variational Monte Carlo with recurrent neural-networks for approximating the ground state of a many-body Hamiltonian. We briefly review the key ingredients of each algorithm and their corresponding neural-network implementation, and show numerical experiments for a system of interacting Rydberg atoms in two dimensions.
\end{abstract}

\maketitle

\section{Introduction}

Quantum many-body physics refers to the mathematical framework to study the collective behavior of large numbers of interacting particles. The emerging cooperative phenomena that result from seemingly simple interactions can produce an astounding variety of phases of matter such as conventional metals and magnetically ordered states, as well as unanticipated states including high-temperature superconductivity, strange metals, and spin liquids~\cite{Xiao:803748}. In addition to naturally occurring quantum systems, many-body physics studies synthetic quantum matter, e.g., ultracold atoms, superconducting qubits, and trapped ions, which simultaneously reveal new phenomena in highly controlled laboratory settings and advances the development of quantum computers and other quantum information processing devices.

In spite of the simplicity of the physical laws that govern such multi-particle quantum objects, the theoretical and experimental analysis of these systems confront us with complexities which are ultimately rooted in the "curse of dimensionality" associated with the exponential explosion of the size of the space where quantum many-body states live in. Traditionally, the study of many-body systems is performed with the help of tools designed to circumvent this dimensionality explosion and produce a succinct, low-dimensional description that captures the essential aspects of a quantum system. Such descriptions arise from the analysis of data generated in a wide range of theoretical, computational, and experimental devices. These include numerical simulations of model Hamiltonians based on quantum Monte Carlo or variational algorithms, but also experimental arrays of complex electronic-structure images obtained from spectroscopic imaging scanning tunnelling microscopy, or measurements of quantum states prepared on a physical quantum computing platform.

Machine learning, already explored as a tool in several research areas in physics~\cite{RevModPhys.91.045002}, offers a set of alternative approaches to the study of quantum many-body systems in experiments and numerical simulations~\cite{doi:10.1080/23746149.2020.1797528,annurev-conmatphys-031119-050651}. The resurgence of activity at the intersection between physics and machine learning is in part due to a series of scientific breakthroughs in computer vision and natural language processing. Such progress has led to a burst of research where neural networks have been repurposed to tackle fundamental questions in condensed matter physics, quantum computing, statistical physics, and atomic, molecular and optical physics. Machine learning, and in particular deep neural networks, have been used to identify phases of matter in numerical simulations and experiments~\cite{carrasquilla2017nature, evert2017nature, torlai_learning_2016, leiwang2016, chng2017, broecker2017, eun-ah2017, dassarma2017, neupeurt2017, yi-ting2018, huembeli2018,PhysRevB.99.060404,PhysRevB.99.104410,Zhang_MLcuprates,Bohrdt2018,Rem2018,PhysRevLett.122.210503}, 
to increase the performance of Monte Carlo simulations~\cite{huang2017,junwei2017, xiao_yan2017, inack2018, parolini2019, pilati2019, mcnaughton2020, albergo2019}, to accurately describe the state of classical~\cite{Wu_2019} and quantum systems~\cite{androsiuk1993,LAGARIS19971, Carleo_2017, zi2018, Di_Luo, pfau2019abinitio, hermann2019deep, PhysRevLett.122.250502, PhysRevLett.122.250501,PhysRevLett.122.250503,PhysRevB.99.214306, choo_fermionicnqs2020, PhysRevLett.124.020503, RNNWF_2020, roth2020iterative}, to develop novel quantum control strategies~\cite{PhysRevX.8.031086,PhysRevX.8.031084,PhysRevLett.122.020601,niu_universal_2019,2020arXiv201003655Y,coopmans2020}, to perform quantum tomography~\cite{torlai_Tomo,rocchetto,Torlai_latent,2018arXiv181206693Q,carrasquilla_povm,biamonte_qst,torlai_rydberg19,xin_local-measurement-based_2019,Sehayek2019,torlai_chemistry,PhysRevA.102.022412,NoriGAN,Tiunov:20,Cha2020,PhysRevA.102.042604,DeVlugt2020,2020arXiv200907601S,torlai_QPT,morawetz2020,Nori2020}, to accelerate density functional theory calculations~\cite{PhysRevLett.108.253002,doi:10.1063/1.4834075,PhysRevB.94.245129,brockherde_bypassing_2017,PhysRevA.100.022512,PhysRevLett.125.076402,PhysRevResearch.2.033388}, to develop and elucidate renormalization group analyses~\cite{2014arXiv1410.3831M,koch-janusz_mutual_2018,PhysRevE.97.053304,PhysRevLett.121.260601,PhysRevResearch.2.023369,2020arXiv201005703C}, to devise quantum error correction protocols~\cite{torlai_neural_2016,krastanov_deep_2017,Varsamopoulos_2017,Baireuther2018machinelearning,Chamberland_2018,Breuckmann2018scalableneural,Nautrup2019optimizingquantum,PhysRevLett.122.200501,PhysRevA.99.052351,Andreasson2019quantumerror,Evert2020QEC,Ni2020neuralnetwork,PhysRevResearch.2.033399,PhysRevResearch.2.023230}, among many other examples~\cite{PhysRevB.97.045153,Seif_2018,Melnikov1221,Dunjko_2018,PhysRevE.99.062106, 2019arXiv191211052C,PhysRevB.99.075113,PhysRevLett.124.010508, PhysRevX.10.011006,2020arXiv200600712H,2020arXiv200905580L,2020arXiv201014510L}. 

Such an explosion of activity indicates that machine learning techniques may soon become commonplace in quantum many-body physics research, both in experiments and numerical simulation. These clear trends call for the development of resources to stimulate researchers to familiarize with the wealth of concepts, intuition, algorithms, hardware, software, and research culture entailed by the adoption of machine learning and neural networks in physics research. Here, we take a step forward in this direction and develop a set of hands-on tutorials focused on a set of recent prototypical examples of applications of neural network technology to problems in statistical physics, condensed matter and quantum computing. 

\subsection*{Outline}
The Article is organized as follows. Starting with a preliminary discussion, we introduce in Sec~\ref{preliminariesA} some fundamental concepts in machine learning and neural networks. In Sec~\ref{preliminariesB} we present a concise description of the physical system studied in our numerical experiments, a two-dimensional array of interacting Rydberg atoms. In Sec~\ref{supervised} we discuss our first application, the classification of phases of matter with supervised machine learning of projective measurement data using a convolutional neural network, and demonstrate it on the quantum phase transition in the Rydberg atoms. In Sec~\ref{qst} we introduce quantum state tomography, and show how this problem can be phrased as an unsupervised machine learning task. Using the restricted Boltzmann machine, we show quantum tomography of the Rydberg ground states, as well as of the ground state of a small molecule from qubit measurement data. In Sec~\ref{vmc} we present the simulation of the ground state of a many-body Hamiltonian using variational Monte Carlo with a recurrent neural network wavefunction. For each of these applications, we also show the key components of the underlying software, with full code tutorials available in an external repository~\cite{coderepo}.

\section{Preliminaries}
\subsection{Machine learning with neural networks}
\label{preliminariesA}
Artificial intelligence, the scientific discipline that deals with the theory and development of computer programs with the ability to perform complex tasks, saw early success solving problems which are relatively straightforward to formalize in an abstract way. The solutions to this breed of problems are typically described by a list of very precise formal rules that computers can process efficiently. As remarkable example, computers have been beating humans at playing chess since 1997, due in part to the fact that chess involves a large set of formal rules.

Modern machine learning, instead, deals with the challenge of automatizing the solution of real world tasks that may be easy for humans to process but that are hard to formally describe by simple rules. These techniques have spurred a recent revolution where algorithms trained using data have  started to match humans' ability to recognize objects in an image, decipher speech or translate text to multiple languages, which are tasks that are difficult to formalize and articulate through simple rules. 

A key element behind these recent developments can be largely traced back to a series of breakthroughs in the development of powerful neural network models, where data is processed through the sequential combination of multiple nonlinear layers~\cite{Goodfellow-et-al-2016}. Such models solve a fundamental problem in learning real world tasks, namely the problem of automatically extracting knowledge from raw noisy data, rather than relying on hard-coded knowledge directly inscribed in the algorithms by a human. Neural networks automatize the construction of sets of increasingly complex representations of the data, which can be understood as the computational disentangling of complex concepts (e.g. an object in a cluttered image) out of simpler concepts (e.g. pixel values and basic shapes like edges). These representations, in turn, lead to solutions to learning tasks with unprecedented success. 

% ML tasks 
\begin{figure*}[t]
\noindent \centering{}\includegraphics[width=2.05\columnwidth]{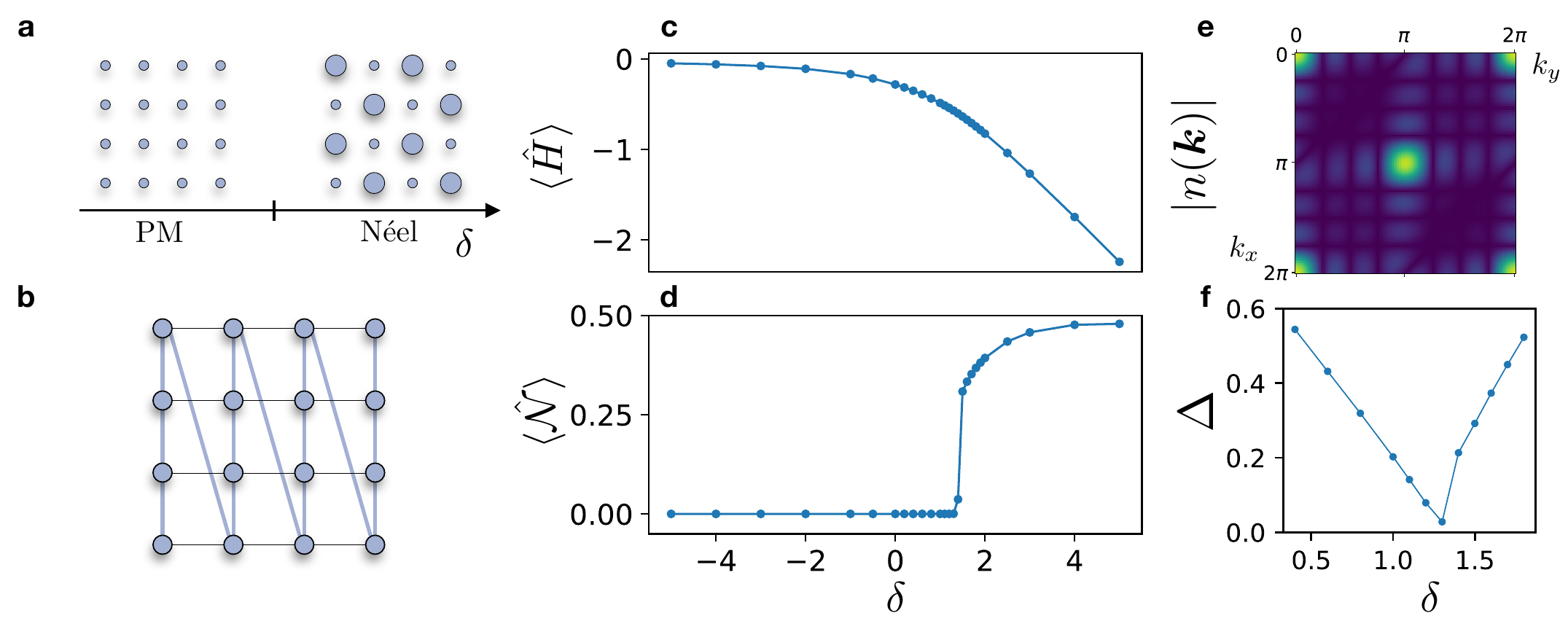}
\caption{Rydberg atoms in a two-dimensional square array. ({\bf a}) Schematic representation of the phase diagram at a fixed value of the interaction $V=3$ MHz and $\Omega = 1$ MHz. At large and negative detuning, the system is in a disordered (paramagnetic) phase with all atoms in the ground state. At large and positive detuning, the atoms are found in a checkerboard pattern with N\'eel order. ({\bf b}) Snake-like geometry of the MPS path along the square lattice, used for the DMRG simulations. Ground state energy ({\bf c}) and the staggered magnetization (N\'eel order) ({\bf d}) as a function of the detuning $\delta$, for a $8\times8$ array ($V=3$ MHz). ({\bf e}) Absolute value of the average occupation number in momentum space $|n(\bm{k})|$ deep into the $Z_2$ ordered phase ($\delta=4$ MHz), showing a peak at $\bm{k}=(\pi,\pi)$, a signature of anti-ferromagnetic order. ({\bf f}) Energy gap $\Delta$ between the ground state and first excited state, detecting a quantum phase transition at detuning $\delta\approx1.3$.}
\label{Fig::1} 
\end{figure*}

For practical purposes, machine learning algorithms can be divided into the categories of supervised, unsupervised, and reinforcement learning, all of which have found applications to quantum many-body systems~\cite{doi:10.1080/23746149.2020.1797528}. While there is no formal difference between some of the algorithms in these categories when expressed in the language of probability~\cite{Goodfellow-et-al-2016,10.5555/1162264}, such a division is often used as a way to specify the details of the algorithms, the training setup, and the structure of the data sets involved. 

Supervised learning tasks aim at predicting a target output vector $\bm y$ associated with input vector $\bm x$,  both of which can be discrete or continuous. The training data is thus a list of pairs of input/output tuples $\{ \bm x_i,\bm y_i\}_{i=1}^{M}$, where target output conveys that such a vector corresponds to the ideal output given the input vector~\cite{10.5555/1162264}. Starting with a training data set with $M$ entries, the learning algorithm outputs a function $\hat{\bm y} = f(\bm x)$  which estimates the output values for unseen input vectors $\bm x$. Examples of supervised learning include classification, where the objective is to assign each input vector to one of a set of discrete categories, and the task of
regression, where the output is a vector with continuous entries. Examples for classification and regression are respectively the problem of recognizing images of handwritten digits and the problem of determining the orbits of bodies around the sun from astronomical data. 

Unsupervised learning deals with the learning tasks where the training data is composed of a set of input vectors without a corresponding target output~\cite{10.5555/1162264}. These algorithms are typically used to discover hidden structure in the data sets.  Examples of tasks in unsupervised learning problems include clustering, where the objective is to discover of groups of similar examples within the data, density estimation, where the objective is to estimate the underlying probability distribution associated with the data, as well as low-dimensional visualization of high-dimensional data algorithms, which depict complex data in two or three dimensions while trying to retain key spatial characteristics in the original data. 

Finally, reinforcement learning, although not discussed in this Article, develops algorithms  dealing with the problem of discovering actions that maximize a numerical reward signal~\cite{10.5555/551283}. The learning algorithms are not necessarily directly exposed to examples of optimal actions. Instead, it must discover them by a process similar to a guided trial and error. Reinforcement learning augmented by deep neural networks has successfully learned policies from high-dimensional sensory input for game playing achieving human-level performance in several challenging games including Atari 2600~\cite{mnih_human-level_2015} as well as the board game Go~\cite{silver2016}. Likewise, reinforcement learning has been applied to the control of quantum systems~\cite{bukov2018,niu_universal_2019} as well as to the optimization of quantum error correction codes~\cite{Nautrup2019optimizingquantum,Andreasson2019quantumerror,Evert2020QEC}, one key ingredient in the development of fault-tolerant quantum computers.

\subsection{Rydberg atoms in two dimensions}
\label{preliminariesB}
We demonstrate the machine learning algorithms discussed in this Article for a many-body system composed by interacting Rydberg atoms. Engineered arrays of cold Rydberg atoms are increasingly used for highly-controlled quantum simulations of strongly-interacting   matter~\cite{Schauss1455,Endres2016,Labuhn,Bernien2017,keesling_quantum_2019,2020arXiv201212281E,2020arXiv201212268S}, as well as for quantum information processing~\cite{Levine2019}. We specifically consider a square array with linear dimension $L$ containing $N = L^2$ atoms. Each atom is described by a local Hilbert space spanned by the states $\{|g\rangle,|e\rangle\}$, referring respectively to the atomic ground and the highly-excited Rydberg states. The atoms are subject to a uniform laser drive with Rabi frequency $\Omega$ and detuning $\delta$, and they interact with one another via the Van der Waals potential $V(x)\approx r^{-6}$ at short distances. The resulting many-body Hamiltonian is
\begin{equation}
\hat{H} = -\Omega \sum_{\bm{r}}\hat{S}^x(\bm{r}) -\delta\sum_{\bm{r}}^N\hat{\Pi}(\bm{r}) +\frac{1}{2}\sum_{\bm{r},\bm{r^\prime}}V(\bm{r}-\bm{r^\prime})\hat{\Pi}(\bm{r}) \hat{\Pi}(\bm{r^\prime}) 
\label{Eq::RydbergHamiltonian}
\end{equation}
where $\hat{\Pi}(\bm{r}) =|e\rangle\!\langle e|_{\bm{r}} $ is the projector onto the Rydberg state at position $\bm{r}$, $\hat{S}^x(\bm{r})=\frac{1}{2}\hat\sigma^x(\bm{r})$ are spin-$\frac{1}{2}$ operators, and $V(\bm{r}-\bm{r^\prime})=V_0/\|\bm{r}-\bm{r^\prime}\|^6$ is the Van der Waals potential between atoms at position $\bm{r}$ and $\bm{r^{\prime}}$. In the following, we assume $\Omega=1$ MHz. 

The phase diagram for the ground state of the Rydberg Hamiltonian is dictated by the mechanism of Rydberg blockade, a constraint that prevents two atoms at sufficiently small distances to be simultaneously excited to the Rydberg states. We can characterize the phase diagram in terms of the detuning $\delta$ and the interaction strength $V_0$. On the square lattice, several different orders have been detected by numerical simulations~\cite{Samajdar2020}. Here, we specifically focus on the $Z_2$ transition between a disordered phase at large and negative detuning, where all atoms are found in the ground state, and an ordered phase at large and positive detuning, where the system is found in one of the two symmetry-broken N\'eel states  characterized by a checkerboard pattern in the atomic occupation number (Fig.~\ref{Fig::1}(a)). 

We perform numerical simulations of the ground state of Hamiltonian~(\ref{Eq::RydbergHamiltonian}) using the density matrix renormalization group (DMRG)~\cite{PhysRevLett.69.2863,PhysRevB.48.10345,SCHOLLWOCK201196} implemented using the software package ITensor~\cite{itensor}. We adopt a matrix product state (MPS) variational wavefunction $|\Psi\rangle$ with a snake-like geometry shown in Fig.~\ref{Fig::1}(b). We fix the interaction strength to $V_0=3$ MHz, and retain up to the third-nearest-neighbor interactions. For a several values of the detuning $\delta\in\{-5,5\}$ MHz, we run DMRG to find an approximation of the ground state, using a singular value decomposition cutoff of $10^{-10}$ and a target energy accuracy of $10^{-5}$. To certify convergence to the ground state, each run is repeated for different initialization of the starting MPS. 

We show the results of the simulations for a $8\times8$ array with open boundary conditions in Fig.~\ref{Fig::1}(c-f). We plot, as a function of the detuning, the ground state energy per site $E_0/N=\langle\Psi_0|\hat{H}|\Psi_0\rangle/N$, and the staggered magnetization  $\langle\mathcal{N}\rangle=N^{-1}\sum_{\bm{r}}(-1)^{x+y}\langle\hat{S}^z(\bm{r})\rangle$, which can be used to detect N\'eel order. Whenever all atoms are in the ground state, $\langle\mathcal{N}\rangle\approx0$, while for an ordered state with a checkerboard pattern one has $\langle\mathcal{N}\rangle\approx0.5$. We also show the average occupation number in momentum space, 
\begin{equation}
n(\bm{k})=\frac{1}{\sqrt{N}}\sum_{\bm{r}}e^{i\bm{k}\cdot\bm{r}}\langle\hat{n}(\bm{r})\rangle
\end{equation}
where $\hat{n}(\bm{r}) = \frac{1}{2}(1-2\hat{S}^z(\bm{r}))$. We observe a peak at $\linebreak\bm{k}=(\pi,\pi)$ for large detuning $\delta=4$ MHz (Fig.~\ref{Fig::1}(e)), and a featureless state at negative detuning (not shown).

The two phases of the Rydberg atoms are separated by a second-order quantum phase transition at a critical point $\delta_c$. We can  extract an approximation of $\delta_c$ by measuring the energy gap $\Delta = |E_0-E_1|$ between the ground state and the first excited state $|\Psi_1\rangle$. We compute $E_1$ by running DMRG on the Hamiltonian $\hat{H}^\prime=\hat{H}+\omega|\Psi\rangle\langle\Psi|$ where $\omega$ is an energy penalty. From the energy gap curve, we estimate the detuning where $\Delta\approx0$ to be $\delta_c\approx 1.3$ MHz. This approximate value will be sufficient for the purpose of this Article, though a more systematic scaling study with appropriate boundary conditions (to minimize finite-size effects) should be performed to accurately determine the critical point and critical exponents of the transition.

Once we have solved for the ground states of the Rydberg Hamiltonian, the corresponding MPSs can be used to generate data to train the neural networks for the different applications. In this case, the data consists of projective measurements in the atomic occupation number basis $|\bm{\sigma}\rangle=|\sigma_1,\dots,\sigma_N\rangle$, where $\sigma_j=0$ and $\sigma_j=1$ refers respectively to the $j$-th atom being in the ground and Rydberg state. Given a wavefunction $|\Psi\rangle$, the probability to observe an atomic pattern $\bm{\sigma}$ following a measurement is simply given by the Born rule $P(\bm{\sigma})=|\langle\bm{\sigma}|\Psi\rangle|^2$. Because of the intrinsic one-dimensional geometry of an MPS, it is possible to efficiently sample the probability distribution $P(\bm{\sigma})$ by exploiting the chain rule of probabilities. Moreover, the sampling is exact in the sense that each sample is completely independent of one another~\cite{PhysRevB.85.165146}. %The sampling procedure consists of iteratively building one-site reduced density matrices conditional on the previous measurement outcomes~\cite{PhysRevB.85.165146}. 

%----------------------------------------------------------------------------------------
%----------------------------------------------------------------------------------------
%                                  SUPERVISED LEARNING
%----------------------------------------------------------------------------------------
%----------------------------------------------------------------------------------------
%----------------------------------------------------------------------------------------

\section{Learning a quantum phase transition}
\label{supervised}

An important task in condensed matter and statistical physics is to characterize different phases of matter and the associated phase transitions between them.
Typically, phases of matter are described in terms of simple real-space patterns and their associated order parameters, which are theoretically understood using Landau symmetry-breaking paradigm~\cite{Xiao:803748}. While a wide array of theoretical and experimental tools to study interacting quantum systems have been constructed in relation to these patterns, there is an increasing set of states of matter whose theoretical and experimental understanding eludes the Landau symmetry-breaking paradigm. The characterization of these phases may rely on, e.g., out-of-equilibrium properties of the system as in the many-body localized phase~\cite{basko2006,RevModPhys.91.021001}, or on topological invariants in topological phases and spin liquids~\cite{Xiao:803748,savaryQuantumSpinLiquids2016, doi:10.1146/annurev-conmatphys-031218-013401}.    

Machine learning provides an alternative route to the characterization of phases of matter and their associated phase transitions in a semi-automated fashion without a direct use of manually specified real-space patterns and/or other signatures, provided that a sufficiently large training set is available. In its simplest form~\cite{carrasquilla2017nature}, given the existence of a classical or quantum phase transition between two phases in a physical system, one can use supervised learning to attempt to classify experimental or numerical snapshots of the phases of matter separated by the transition. This task can be achieved using most classification algorithms, e.g., those based on a neural network or a support vector machine~\cite{10.5555/1162264}, trained on snapshots of two phases of matter labelled according to the corresponding phase out of which the snapshot originated. Although here we only explore this simple strategy, we stress that machine learning approaches to studying phases and phase transitions have been significantly expanded and they no longer require the precise knowledge of the location of the critical point~\cite{evert2017nature,broecker2017b}, can be fully automatized, and can discover ordered phases~\cite{carrasquilla2017nature,leiwang2016,PhysRevE.96.022140}, topological phases~\cite{evert2017nature,rodriguez-nieva2019,PhysRevLett.125.127401}, and phases such as the many-body localized phase which is characterized by its dynamical properties~\cite{yi-ting2018,rao2018,PhysRevLett.120.257204}.  

The nature of the snapshots used to train the learning algorithms is vastly flexible, hence these strategies are of wide applicability, and can include numerically generated configurations visited during a classical or quantum Monte Carlo simulation of the physical system~\cite{carrasquilla2017nature,leiwang2016,evert2017nature,broecker2017,chng2017,PhysRevB.99.060404,PhysRevB.99.121104}, entanglement spectra~\cite{evert2017nature,yi-ting2018}, correlation matrices~\cite{PhysRevB.102.054512,PhysRevLett.125.170603}, tensors in an MPS~\cite{PhysRevLett.125.170603}, numerically generated projective measurements~\cite{berezutskii2020}, high-resolution real-space snapshots of complex many-body systems obtained with quantum gas microscopes for ultracold atoms~\cite{Bohrdt2018,PhysRevA.102.033326}, single-shot experimental momentum-space density images of ultracold quantum gases~\cite{Rem2018}, spectroscopic imaging scanning tunnelling microscopy data~\cite{Zhang_MLcuprates}, among many others. 

Below we explore learning a quantum phase in an array of interacting Rydberg atoms using projective measurements. As a classification algorithm, we make use a of a convolutional neural network~\cite{Goodfellow-et-al-2016} that readily takes advantage of the two-dimensional (2D) spatial arrangement of the the Rydberg atoms and their locality, as well as the approximate translation invariance of the system. 

\begin{figure}[t]
\noindent \centering{}\includegraphics[width=\columnwidth]{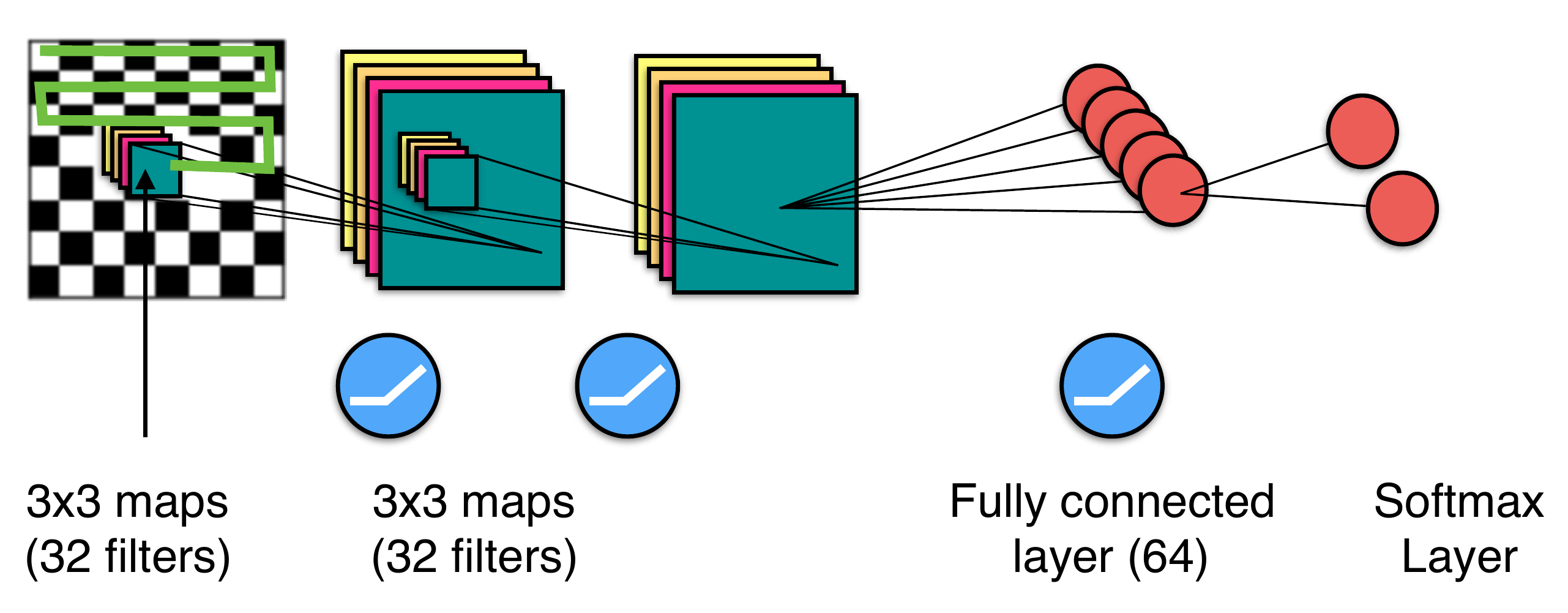}
\caption{A schematic representation of a convolutional neural network. The elements of the input $\mathsf{h}^{(0)}_{l,j,k}$ corresponds to the outcome of a projective measurement on the Rydberg system. The first operation is a convolutional layer with $ M_y \times M_x = 3  \times 3$ kernels  with $I_{\text{out}} = 32$ output channels and $L_{\text{input}}=1$. This kernel is convolved with an input configuration with $N = 8 \times 8$ Rydberg atoms. Likewise, the second operation corresponds to a convolutional layer with $ M_y \times M_x = 3  \times 3$ kernels  with $I_{\text{out}} = 32$ output channels and $L_{\text{input}}=32$. The output of the second convolutional layer is flattened and fed to an FC layer with a ReLU activation, followed by another FC layer with a softmax activation which produces the prediction outcome. }
\label{Fig::CNN} 
\end{figure}

\subsection{Convolutional neural networks and their training}
Convolutional neural networks (CNN) employ a mathematical operation called convolution to process information for data that has a natural grid-like topology~\cite{Goodfellow-et-al-2016}. A 2D convolutional layer implements the operation
\begin{align*}
\mathsf{h}^{(q)}_{i,j,k}  & =  F\left(\sum_{l,m_y,m_x} \mathsf{h}^{(q-1)}_{l,j+m_y,k+m_x} \mathsf{K}^{(q)}_{i,l,m_y,m_x} \right)  \\
&\coloneqq   F\left( \mathsf{K}^{(q)} *\mathsf{h}^{(q-1)}  \right) \nonumber %_{i,j,k} \nonumber
\end{align*}
where the trainable kernel $\mathsf{K}^{(q)}_{i,l,m_y,m_x}$ at layer $q$  specifies the connection strength between a unit in channel $i$ of the output and a unit in channel $l$ of the input, with a spatial offsets of $m_y$ rows (labeled y direction) and $m_x$ columns (labeled x direction) between the output and the input variables. The dimensions of the array $\mathsf{K}^{(q)}_{i,l,m,n}$ are $I_{\text{out}}$, $L_{\text{input}}$, $M_{y}$, $M_{x}$, which corresponds to the number of output channels, input channels, dimension of the filter along the vertical and horizontal directions, respectively. The activation at layer $q$ consists of  elements $\mathsf{h}^{(q)}_{l,j,k}$, where $j$ and $k$ label vertical and horizontal directions, respectively, and $l$ specifies the channel. The activation units are labelled by $q$ where $q=0$ corresponds to the raw projective measurement data. Finally, the non-linear function $F(x)$, which in our examples is typically a rectified linear unit (ReLU) $F(x)=\text{max}(0,x)$, is applied element-wise to each of the components of its input. A convolutional neural network equipped with two convolutional layers is schematically shown in Fig.~\ref{Fig::CNN}.

Followed by the convolutional layers, a CNN typically processes information using sets of fully connected (FC) layers which implement a matrix-vector operation followed by a non-linearity $F$ as
\begin{equation}
\mathsf{h}^{(q)}_{i}  =  F\left(\sum_{l} \mathsf{h}^{(q-1)}_{l}  \mathsf{K}^{(q)}_{i,l} + \mathsf{b}^{(q)}_{i}\right), 
\end{equation}
where the trainable parameters of the FC layer are the kernel $\mathsf{K}^{(q)}_{i,l}$ and the bias vector $ \mathsf{b}^{(q)}_{i}$. 
To feed  the output of a convolutional layer $\mathsf{h}^{(q-1)}_{l,j,k}$ to an FC layer, the array is reshaped or ``flattened'' to   $\mathsf{h^{\prime}}^{(q-1)}_{l}$ so that all the original components packed into a one-dimensional array with dimension $L_{\text{FC}}$. The last two layers of the CNN in Fig.~\ref{Fig::CNN} correspond to two fully connected layers with a ReLU  and a softmax non-linearities, respectively. The softmax function $S$ is given by 
\begin{equation}
  \text{S}(\bm{v}) = \frac{\exp(\bm{v})}{\sum_i \exp(v_i)}.  
\end{equation}
where $v_i$ are the components of a vector $\bm{v}$ and the $\text{exp}$ function acts element-wise on the components of the vector. We note that the input to the CNN and its trainable parameters are real, so that the outcome of the softmax layer can be interpreted as a probability distribution since $0 \le \text{S}(v_i)\le 1$ and $\sum_{i}\text{S}(v_i)=1$.

Finally, we mention that we interpret our CNN as a model for the conditional probability of assigning a phase of matter $y=0,1$ to a projective  measurement outcome $\bm{\sigma}=\mathsf{h}^{(0)}$, i.e. $P_{\bm{\theta}}(y|\bm{\sigma})$, where $\bm{\theta}$ encompasses all the trainable parameters of the CNN. The conditional is given by 
\begin{widetext}
\begin{equation}
 P_{\bm{\theta}}(y|\bm{\sigma}) = S\left( b^{(4)}_{y} + \sum_{m}\mathsf{K}^{(4)}_{y,m} F\left(b^{(3)}_{m} + \sum_{l} \mathsf{K}^{(3)}_{m,l}\,\text{Flatten}\left(F\left((\mathsf{K}^{(2)}*F\left(\mathsf{K}^{(1)}*\bm{\sigma} \right)\right)\right)_{l}\right)       \right), 
 \label{Eq::CNNdist}
\end{equation}
\end{widetext}
where the function $\text{Flatten}()_{l}$ is the $l$-th component of a vector that arises from reshaping the incoming argument of the function to a one-dimensional array.  

%\subsection{Phase transition in the Rydberg array}

To estimate the parameters of the CNN we use the maximum likelihood principle, where the parameters of a statistical model are selected by assigning high probability to the observed data. For a dataset with observations $\{\bm{\sigma}_n, y_n \}_{n=1}^{M}$, where $y_n=0,1$ label the phase of matter out of which a projective measurement $\bm{\sigma}_n$ was taken from, the likelihood assigned by the model to the dataset can be written as 
\begin{equation}
    p(\bm{y}|\bm{\theta}) = \prod_{n=1}^{M}
     P_{\bm{\theta}}(y_n|\bm{\sigma}_n)^{y_n}(1-P_{\bm{\theta}}(y_n|\bm{\sigma}_n))^{1-y_n}
\end{equation}
where $\bm{y}=(y_1,...,y_{M})$. Instead of attempting to maximize the likelihood, it is convenient to define a loss function by taking the negative logarithm of the likelihood, which gives the cross-entropy
\begin{widetext}
\begin{equation}
E(\bm{\theta}) = -\ln( p(\bm{y}|\bm{\theta})) - =  \sum_{n=1}^{M} \{ y_n \ln \left(P_{\bm{\theta}}(y_n|\bm{\sigma}_n))\right) + (1-y_n)\ln(1 - P_{\bm{\theta}}(y_n|\bm{\sigma}_n)) \}. 
\label{Eq::NLL}
\end{equation}
\end{widetext}

To train the model, we minimize $E(\bm{\theta})$ using gradient descent techniques~\cite{10.5555/1162264}. While it is possible to evaluate the gradients of $E(\bm{\theta})$ with respect to the parameters $\bm{\theta}$ in the CNN analytically using the chain rule, a more convenient and less error-prone approach is to use automatic differentiation (AD), which is a set of techniques to numerically evaluate the derivative of a function specified by a computer program. A complete survery detailing AD can be found in Ref.~\cite{AD_2017}. 

In addition, instead of using the entire dataset in the calculation of $E(\bm{\theta})$ and its gradients, we use smaller batches of data of size $M_{\text{batch}}<M$, which means that the gradients used during optimization become stochastic since they fluctuate from batch to batch. In the examples below we use $N_{\text{batch}}=32$. The gradient update rule used in our examples is a modified version of the usual gradient descent called Adam~\cite{2014arXiv1412.6980K}.

\subsubsection*{Code walkthrough}
We demonstrate supervised learning with a CNN to learn the quantum phase transition in the Rydberg array, using the machine learning software library TensorFlow~\cite{tensorflow}. We first generate training data by sampling the MPS wavefunctions obtained from DMRG at different detunings $\delta$. This data is then divided into a training set (used to update the neural network parameters) and a test set (used to validate the performance of the model). Each data set consists of a list of atomic occupation patterns $\bm\sigma$ and their ``phase label'' $y$.

We begin by importing the required functionalities and loading the data. Since we are using a CNN with a two-dimensional geometry, the atomic configurations in the training and test data sets need to be appropriately reshaped from the one-dimensional MPS structure.
\begin{lstlisting}[language=Python,numbers=none]
import numpy as np
import tensorflow as tf
from tensorflow.keras import layers, models

# linear dimensions of the system
Lx = 8
Ly = 8

# training set
train_config = np.loadtxt("xtrain.txt") 
train_label = np.loadtxt("ytrain.txt",dtype=np.uint8)

# test set
test_config = np.loadtxt("xtest.txt")
test_label  = np.loadtxt("ytest.txt",dtype=np.uint8)

# reshaping the training/test configurations
train_config = np.reshape(train_config,(train_config.shape[0],Lx,Ly,1))
test_config = np.reshape(test_config,(test_config.shape[0],Lx,Ly,1))

# reshaping the training/test labels
test_label  = np.reshape(test_label,(test_label.shape[0],1))
train_label = np.reshape(train_label,(train_label.shape[0],1))

# Names for the phases  
class_names = ["disordered", "ordered"]
\end{lstlisting}

Next, we define the neural network architecture by combining layers pre-defined in TensorFlow. After initialization, we proceed to implement the model of Eq.~\ref{Eq::CNNdist}. First,  the raw input data is processed by two stacked convolutional layers, each one with $3\times3$ filters and 32 channels using rectified linear units. The output of the second CNN layer is fed to two stacked fully-connected layers with $64$ hidden units, and two output units. These last units correspond to model output for the disordered and ordered phase. 
\begin{lstlisting}[language=Python,numbers=none]
# initialize the model
model = models.Sequential()

# first convolutional layer
model.add(layers.Conv2D(32, (3, 3),
          activation = "relu", 
          input_shape = (Lx, Ly, 1)))
          
# second convolutional layer
model.add(layers.Conv2D(32, (3, 3), 
          activation="relu"))

# flatten the output
model.add(layers.Flatten())

# dense layer with 64 output units
model.add(layers.Dense(64, activation='relu'))

# dense layer with two output units
model.add(layers.Dense(2))
\end{lstlisting}

Once the architecture is defined, the model can be compiled by adding the cost function (i.e. the cross-entropy in Eq.~\ref{Eq::NLL} ), and the optimizer to update the model parameters, the Adam optimizer~\cite{2014arXiv1412.6980K}. Internally, TensorFlow builds a computational graph containing each operation being executed from the input state to the final output of the architecture. The gradients of the cost function with respect to each network parameter are then evaluated using AD. At compilation time, we can also add a metric to be monitored during training, in this case being the classification accuracy, i.e. the fraction of the test set samples that are being classified correctly.
\begin{lstlisting}[language=Python,numbers=none]
# Compiling the model
model.compile(optimizer='adam', 
  loss=tf.keras.losses.SparseCategoricalCrossentropy(from_logits=True), 
  metrics=['accuracy'])
\end{lstlisting}

The model is now ready to be trained using the Rydberg data for a set number of epochs, which is the number of passes of the entire training data set the machine learning algorithm has completed. After training is complete, we can evaluate the model on the held-out test data to quantify its accuracy.
\begin{lstlisting}[language=Python,numbers=none]
# training the model
history = model.fit(
  train_config, 
  train_label, 
  epochs=5, 
  validation_data = (test_config, test_label))

# evaluate the overall model
test_loss,test_acc = model.evaluate(test_config, test_label, verbose=2)
\end{lstlisting}

\begin{figure}[t]
\noindent \centering{}\includegraphics[width=\columnwidth]{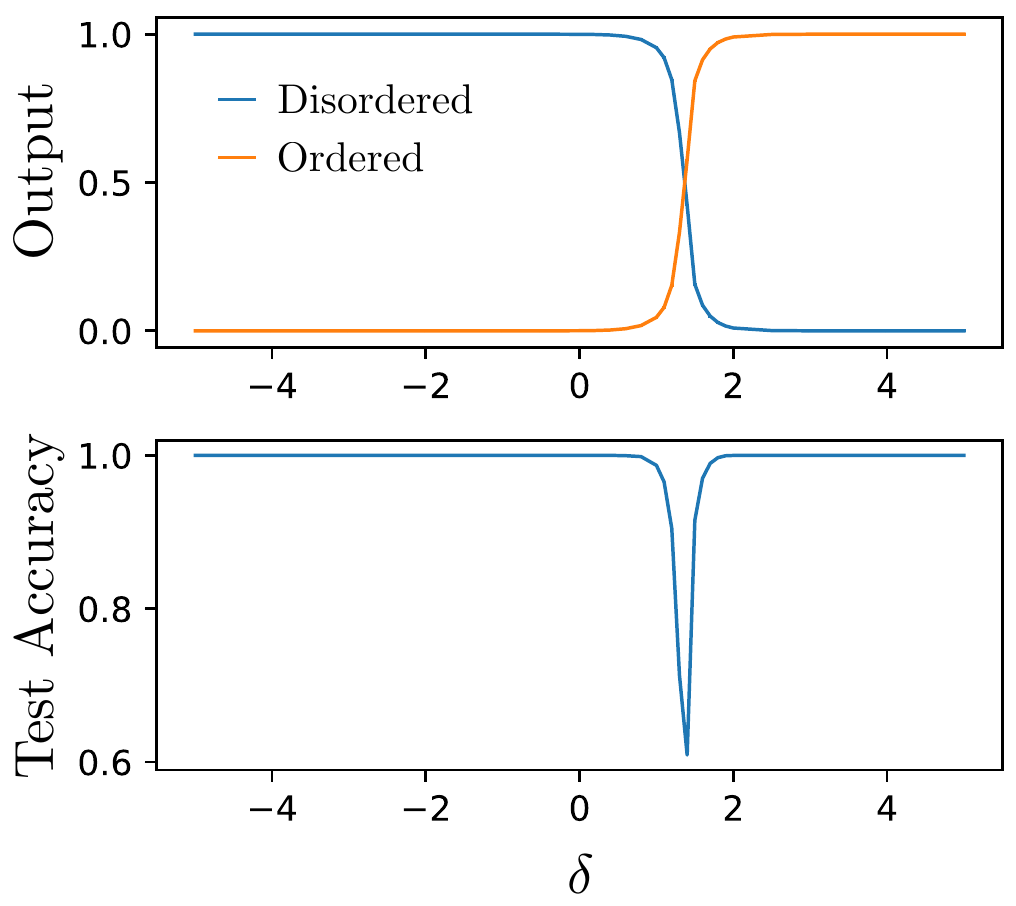}
\caption{Learning the quantum phase transition in the Rydberg-atom array. ({\bf a}) Output signal for the two units in the top-most dense layer of the neural-network architecture as a function of the detuning. ({\bf b}) Accuracy on the test data set, showing a dip near the critical point. This signal can be used to detect the quantum phase transition.}
\label{Fig::supervised} 
\end{figure}

We show the results of the training in Fig.~\ref{Fig::supervised}, evaluated on various test sets at different detuning values $\delta$. We plot the average output signal for the two output neurons, i.e. an estimate of $f(y,\delta) = \sum_{\bm \sigma} |\Psi(\bm{\sigma})|^2 P_{\bm{\theta}}(y|\bm{\sigma})$, on the top-most dense layer. When the detuning is large and negative, the disordered neuron saturates to one, and the ordered neuron is nearly zero, while the signals reverse at large and positive detuning. We can use the crossing point between the two curves to detect the critical point. We also show the accuracy in the test set, which shows a dip near the critical point. This is when the neural network is most uncertain about assigning a phase label to any given atomic configuration.

%----------------------------------------------------------------------------------------
%----------------------------------------------------------------------------------------
%                               QUANTUM STATE TOMOGRAPHY
%----------------------------------------------------------------------------------------
%----------------------------------------------------------------------------------------
%----------------------------------------------------------------------------------------

\section{Quantum state tomography}
\label{qst}
Quantum characterization, verification and validation is a framework for algorithms and routines used to assess the quality and the performance of experimental quantum hardware and characterize its components~\cite{eisert_quantum_2020}. The workflow underlying these algorithms is inherently data-driven: appropriate measurement data is first collected from the quantum device under examination, and then processed by an algorithm running on a classical computer. Depending on the degree of complexity of the algorithm, different amounts of information can be gained. This could be a single figure of merit, such as the average error rate for a set of quantum gates~\cite{PhysRevA.77.012307,PhysRevA.85.042311,PhysRevA.99.052350}, or the fidelity (or a proxy thereof) between a quantum state prepared by the hardware and a desired reference state~\cite{PhysRevLett.106.230501,aolita_reliable_2015,PhysRevLett.120.190501}. One may be also interested in retrieving the full quantum state generated by a device. This procedure -- the reconstruction of an unknown quantum state from measurement data -- is called {\it quantum state tomography} (QST)~\cite{vogel89,PhysRevA.55.R1561,PhysRevA.63.040303,PhysRevA.64.052312,Jezek2003,Blume_Kohout_2010,PhysRevLett.108.070502,Granade_2017}.
%, and it is a fundamental ingredient of any infra-structure for quantum benchmarking and hardware calibration.

There are two assumptions in QST: the ability to prepare many identical copies of the quantum state $\bm{\varrho}$ of interest, and to repeatedly perform measurements on it. The set of measurements is in general described by positive-operator value measures (POVM)  $\linebreak\bm{\mathcal{M}}=\{\Pi_k\}$~\cite{nielsen_chuang_2010}. The set $\bm{\mathcal{M}}$ is said to be {\it informationally complete} (IC) if it spans the full Hilbert space. An example of informationally (over)-complete measurements are Von Neumann measurements in the Pauli bases, where for a single qubit $\bm{\mathcal{M}}$ contains the six rank-1 projectors into the eigenstates of the Pauli matrices. For an IC set, any quantum state can be uniquely identified by the probabilities of the measurements in $\bm{\mathcal{M}}$, as specified by the Born rule $p(k) = \tr[\bm{\varrho}\:\Pi_k]$.

The simplest method to perform QST is {\it linear inversion}, which reconstructs the quantum state simply by inverting the Born rule using an empirical approximation of the measurement probabilities. One issue of linear inversion is that the reconstructed density operator $\bm{\rho}$ is not necessarily positive, although negative eigenvalues can be appropriately removed to produce a positive state that is closest to the output of linear inversion~\cite{PhysRevLett.108.070502}. A more powerful approach, but also more computationally intensive, is {\it maximum likelihood estimation}~\cite{PhysRevA.55.R1561,PhysRevA.63.040303,PhysRevA.64.052312}, where the state $\bm{\rho}$ is found by minimizing the likelihood function for the observed data under the constraint $\bm{\rho}\ge0$. 

Traditional QST algorithms based on linear inversion or maximum likelihood suffer a complexity that scales exponentially with the number of qubits or particles involved. This exponential scaling stems from two reasons. First, the representation of the quantum state $\bm{\rho}$, which is inevitably exponential in the system size. Second, the {\it sample complexity}, that is the number of measurements that needs to be collected in an experiment to achieve a faithful reconstruction of the quantum state. Typically, statistics from a IC set is required in order to get a good fit, and the size thereof scales exponentially with the number of qubits. For these reasons, traditional QST has remained limited to quantum systems containing only a small number of particles~\cite{haffner_scalable_2005}. 

Several algorithms to overcome this severe complexity have been proposed over the last decade. Notable examples are compressed sensing tomography~\cite{PhysRevLett.105.150401,Flammia_2012,PhysRevLett.106.100401,riofrio_experimental_2017}, permutationally invariant tomography~\cite{PhysRevLett.105.250403,Moroder_2012} and tensor-network tomography~\cite{cramer2009efficient,MPOtomo,Lanyon2017,LeiWang2020}, which rely respectively on the sparsity, translational invariance and low-entanglement of the target quantum state. More recently, a new framework built on neural networks and unsupervised learning has been put forward~\cite{torlai_Tomo}, based on the assumption that most physical states of interest typically contains some degree of {\it structure} (i.e. correlations, symmetries, etc), in the sense that they can be described using a reduced number of parameters (much smaller than the dimension of the Hilbert space). The general idea is to leverage the capability of unsupervised machine learning to autonomously identify such structure in raw data, and compress it using a neural-network representation of the quantum state. 

In what follows, we will focus on pure quantum states, and discuss the extension to mixed states at the end of this Section. We consider a system of $N$ qubits (or any other two-level system) described by a wavefunction $|\Phi\rangle$ with amplitudes $\Phi(\bm{\sigma}) = \langle\bm{\sigma}|\Phi\rangle$ in an appropriate reference basis $|\bm{\sigma}\rangle=|\sigma_1,\dots,\sigma_N\rangle$ ($\sigma_j\in\{0,1\}$). In order to circumvent the scalability issue of standard QST, we adopt a compact representation of a wavefunction expressed in terms of a neural network~\cite{Carleo_2017}. The resulting {\it neural-network wavefunction} is simply a highly non-linear parametric function of the basis states $\psi_{\bm{\theta}}(\bm{\sigma})$, where $\bm\theta$ is a set of parameters (e.g. weights and biases). Several types of neural networks have been successfully implemented to perform QST, including feed-forward neural networks~\cite{biamonte_qst,xin_local-measurement-based_2019}, variational auto-encoders~\cite{rocchetto}, generative adversarial networks~\cite{NoriGAN}, recurrent neural networks~\cite{carrasquilla_povm,morawetz2020} and transformers~\cite{Cha2020}. Here, we will examine the restricted Boltzmann machine.

\subsection{The restricted Boltzmann machine}
The restricted Boltzmann machine is an energy-based model introduced in the early 1980s for generative modeling~\cite{Ackley85,Smolensky1986}, built on a connection between cognitive science and statistical mechanics~\cite{Little74,Little78,Hopfield82}. The RBM features two layers of stochastic binary units: a visible layer $\bm{\sigma}=(\sigma_1,\sigma_2,\dots)$ and a hidden layer $\linebreak\bm{h}=(h_1,h_2,\dots)$, containing respectively $N$ and $n_h$ neurons (or units). The two layers in the RBM are fully connected by a symmetric weight matrix $\bm{W}$, with no intra-layer connections (hence its {\it restricted} nature). The visible units are used to represent the variables relevant to the specific problem at hand, such as the pixel values in an image (or the computational basis states for qubits). The size of the hidden layer is a natural control parameter for the representational power of the model. Note that, since RBMs are universal function approximators~\cite{LeRoux2008}, they can capture any discrete distribution provided the number of hidden units is sufficiently large (possibly exponential in the number of visible units).

The RBM associates to each configuration of the visible and hidden layer $(\bm \sigma,\bm h)$ the energy
\begin{equation}
E_{\bm{\theta}}(\bm{\sigma},\bm{h})=-\sum_{j}\sum_{i}W_{ij}h_i\sigma_j-\sum_{j}b_j\sigma_j-\sum_{i}c_ih_i\:,
\end{equation}
where $\bm\theta=(\bm{W},\bm{b},\bm{c})$ is the set of parameters, and we also introduced biases $\bm b$ and $\bm c$ for the visible and hidden units respectively. Given this energy functional, the (stochastic) RBM units are distributed according to the Boltzmann distribution at temperature $\beta=1$
\begin{equation}
p_{\bm{\theta}}(\bm{\sigma},\bm{h})=Z_{\bm{\theta}}^{-1}e^{-E_{\bm{\theta}}(\bm{\sigma},\bm{h})}\:,
\end{equation}
where the partition function is
\begin{equation}
Z_{\bm{\theta}}=\sum_{\bm{\sigma},\bm{h}}e^{-E_{\bm{\theta}}(\bm{\sigma}_,\bm{h})}\:.
\end{equation}
Importantly, because the network architecture is restricted, we can trace out the latent space explicitly, obtaining the marginal probability distribution over the visible space
\begin{equation}
p_{\bm{\theta}}(\bm{\sigma})=\sum_{\bm{h}}p_{\bm{\theta}}(\bm{\sigma},\bm{h}) = Z^{-1}_{\bm{\lambda}}e^{\mathcal{E}_{\bm\theta}(\bm\sigma)}\:,
\label{Eq::RBM}
\end{equation}
where we defined an ``effective energy''
\begin{equation}
\mathcal{E}_{\bm\theta}(\bm\sigma)=\sum_jb_j\sigma_j+\sum_{i}\left(1+e^{\:\sum_{j}W_{ij}\sigma_j+c_i}\right)\:.
\end{equation}

The main purpose of the RBM is generative modeling, which is the task of learning a representation of an unknown probability distribution from data, allowing the neural network to produce new data points. In other words, the RBM training attempts to discover low-dimensional features in the data to allow generalization beyond the finite-size data set. 

Let us consider a data set $\mathcal{D}=\{\bm{\sigma}_k\}$ with underlying (unknown) probability distribution $q(\bm \sigma)$. The RBM can be trained using unsupervised learning to minimize the distance between the two distributions. Such distance measure is typically expressed in terms of the Kullback-Leibler divergence~\cite{Kullback:1951aa}
\begin{equation}
\text{KL}(q\:|\:p_{\bm{\theta}}) = \sum_{\bm{\sigma}}q(\bm{\sigma})\log\frac{q(\bm{\sigma})}{p_{\bm{\theta}}(\bm{\sigma})}\:,
\end{equation}
with $\text{KL}(q\:|\:p_{\bm{\theta}})>0\:\:\forall q,p_{\bm{\theta}}$ and $\text{KL}(q\:|\:p_{\bm{\theta}})=0$ iff $p_{\bm{\theta}}=q$. 

The exponentially large sum over the full configuration space is approximated using the available data, leading to the cost function
\begin{equation}
\mathcal{C}(\bm{\theta})=
-\frac{1}{|\mathcal{D}|}\sum_{\bm{\sigma}\in\mathcal{D}}\log p_{\bm{\theta}}(\bm{\sigma})-H_{\mathcal{D}}\:,
\label{Eq::KL}
\end{equation}
where $|\mathcal{D}|$ is the size of the data set. Note that, up to a constant data set entropy term $H_{\mathcal{D}}$, the KL divergence simply reduces to the negative logarithm of the likelihood function $\mathcal{L}(\mathcal{D}\:|\:p_{\bm\theta})$.

The RBM can be trained using one of the many flavors of gradient descent. It is straightforward to show that the gradients of the cost function are given by\begin{equation}
\nabla_{\bm\theta}\mathcal{C}({\bm\theta}) = \langle\nabla_{\bm\theta} \mathcal{E}_{\bm\theta}(\bm\sigma)\rangle_{p_{\bm\theta}} -  \langle\nabla_{\bm\theta} \mathcal{E}_{\bm\theta}(\bm\sigma)\rangle_{\mathcal{D}}
\end{equation}
where the gradients $\nabla_{\bm\theta}\mathcal{E}_{\bm\theta}(\bm\sigma)$ can be computed exactly for any sample $\bm{\sigma}$. We see that the gradients $\nabla_{\bm\theta}\mathcal{C}({\bm\theta})$ have two components. First, there is the average over the data points $\langle\nabla_{\bm\theta} \mathcal{E}_{\bm\theta}(\bm\sigma)\rangle_{\mathcal{D}}$, which is fast to compute. In contrast, the average over the model distribution
\begin{equation}
\Big\langle\frac{\partial \mathcal{E}_{\bm\theta}(\bm\sigma)}{\partial \bm\theta}\Big\rangle_{p_{\bm\theta}}=
\frac{1}{Z_{\bm\theta}}\sum_{\bm\sigma}e^{\mathcal{E}_{\bm\theta}(\bm\sigma)}\nabla_{\bm\theta}\mathcal{E}_{\bm\theta}(\bm\sigma)
\end{equation}
requires the partition function, whose calculation is in general intractable. However, this expectation value can be approximated using Monte Carlo, by drawing $N_S$ samples from the model distribution $\{\bm\sigma_i\}\sim p_{\bm\theta}(\bm\sigma)$:
\begin{equation}
\Big\langle\frac{\partial \mathcal{E}_{\bm\theta}(\bm\sigma)}{\partial \bm\theta}\Big\rangle_{p_{\bm\theta}}\approx
\frac{1}{N_S}\sum_{i=1}^{N_S}\nabla_{\bm\theta}\mathcal{E}_{\bm\theta}(\bm\sigma_i)\:.
\end{equation}
This is the most computationally intensive step of the training, and depending on the specific distribution to be learned, advanced Monte Carlo algorithms may be required to collect sufficiently uncorrelated samples.

\subsection{Reconstruction of Rydberg atoms}
Now that we have introduced the main features of the RBM and its training, we are ready to explore its use for QST. As a first application, we examine the reconstruction of Rydberg-atom wavefunctions. An important property of the ground state wavefunction $|\Phi\rangle$ of the Rydberg Hamiltonian (\ref{Eq::RydbergHamiltonian}) is that it is positive in the occupation number basis $\Phi(\bm\sigma)\ge0$ (where $\sigma_j=0$ and $\sigma_j=1$ refers to the ground and excited states). This property follows directly from the representation of the Hamiltonian in this basis, in which all of its off-diagonal elements can be gauged to be negative (i.e. the Hamiltonian is {\it stoquastic} in this basis~\cite{Bravyi2006}). 

The positivity of the target state implies that we may parametrize the neural-network wavefunction simply as $\psi_{\bm\theta}(\bm\sigma) = \sqrt{p_{\bm\theta}(\bm\sigma)}$, for any normalized probability distribution $p_{\bm\theta}(\bm\sigma)$. Here we choose the RBM probability distribution (Eq.~\ref{Eq::RBM}), where the visible units correspond to the atomic occupations. Moreover, because the wavefunction is positive, measurement data from a single basis is sufficient to characterize the state. This means that the QST problem, under these assumptions, is equivalent to unsupervised learning of projective measurement data in the atomic occupation number basis. The tomographic reconstruction of the quantum state is carried out by iteratively changing the RBM parameters to minimize the cost function
\begin{equation}
\mathcal{C}(\bm{\theta})=
%-\frac{1}{|\mathcal{D}|}\sum_{\bm{\sigma}\in\mathcal{D}}\log |\langle \bm \sigma|\psi_{\bm\theta}\rangle|^2=
-\frac{1}{|\mathcal{D}|}\sum_{\bm{\sigma}\in\mathcal{D}}\log p_{\bm\theta}(\bm\sigma)\:.
\label{Eq::KL}
\end{equation}

Upon reaching convergence in the training, the learned RBM can be used to estimate various properties of interest. For a generic observable $\hat{\mathcal{O}}$, the expectation value on the RBM wavefunction reduces to an average over the RBM distribution
\begin{equation}
\langle\hat{\mathcal{O}}\rangle = \frac{\langle\psi_{\bm\theta}|\hat{\mathcal{O}}|\psi_{\bm\theta}\rangle}{\langle\psi_{\bm\theta}|\psi_{\bm\theta}\rangle} =
\frac{1}{Z_{\bm\theta}}\sum_{\bm{\sigma}}p_{\bm\theta}(\bm\sigma)\mathcal{O}_{loc}(\bm\sigma)\:,
\label{Eq::avg_obs}
\end{equation}
where we introduced the so-called ``local observable''
\begin{equation}
\mathcal{O}_{loc}(\bm\sigma) = \frac{\langle\bm\sigma|\hat{\mathcal{O}}|\psi_{\bm\theta}\rangle}{\langle\bm\sigma|\psi_{\bm\theta}\rangle}\:.
\end{equation}
The expectation value in Eq.~\ref{Eq::avg_obs} can then be approximated with a Markov chain using Monte Carlo sampling, similarly to the evaluation of the gradients. We point out that the evaluation of the local observable $\mathcal{O}_{loc}(\bm\sigma)$ remains efficient as long as the matrix representation of $\mathcal{O}$ in the reference basis is sufficiently sparse. This measurement procedure is also useful for monitoring different observables during training, such as average densities and correlation functions. These metrics can be used to assess convergence, since in general the calculation of the KL divergence is intractable as it also requires estimating the partition function $Z_{\bm\theta}$.

\begin{figure}[t]
\noindent \centering{}\includegraphics[width=\columnwidth]{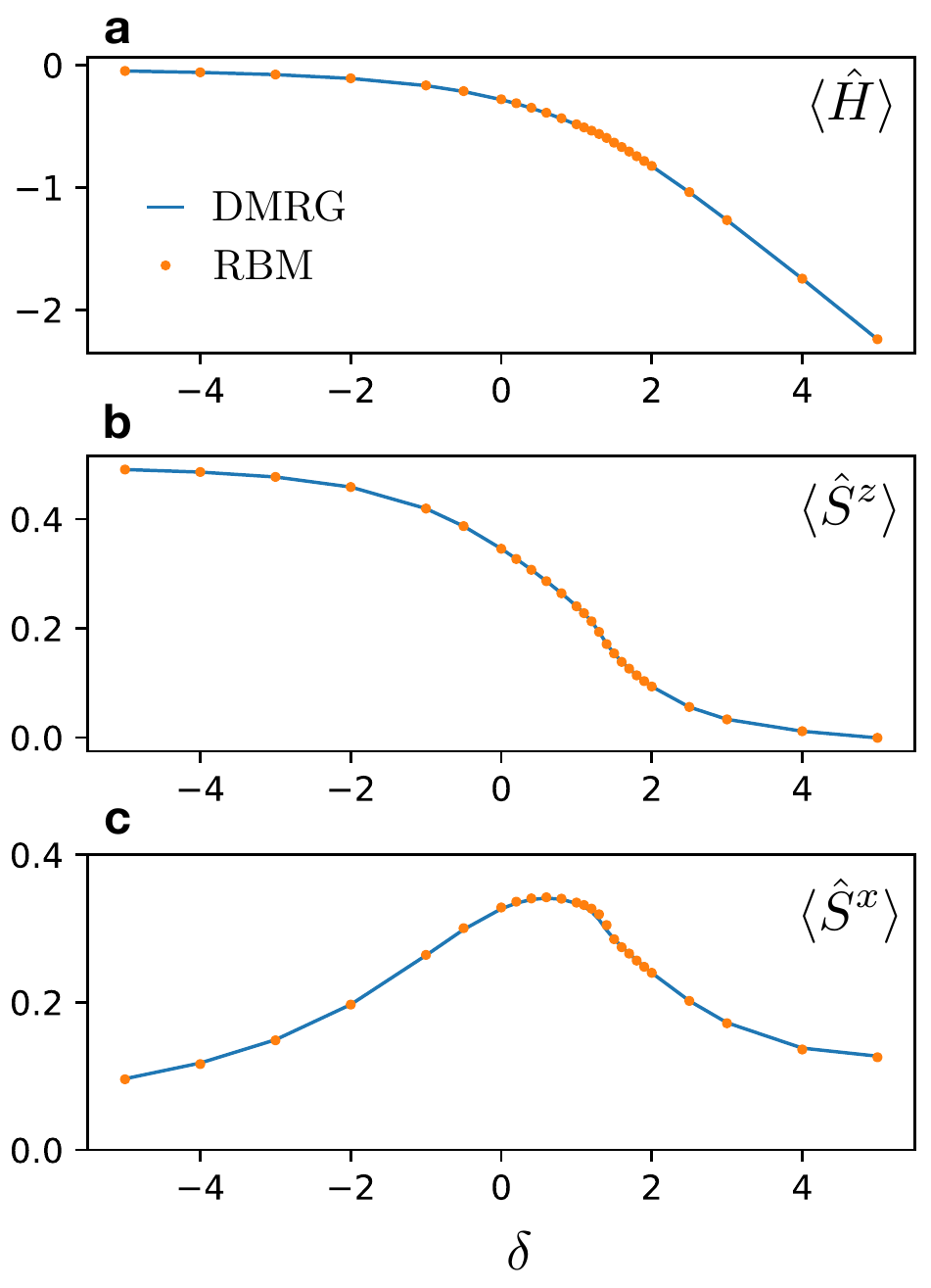}
\caption{Quantum state tomography of a $8\times8$ array of Rydberg atoms with unsupervised learning of single-shot atomic occupation data. We compare various observables measured using the MPSs obtained from DMRG (blue line) and the neural-network wavefunctions at different values of the detuning $\delta$. We plot the average energy per spin ({\bf a}), and the average magnetization along the $z$ ({\bf b}) and $x$ ({\bf c}) axes.}
\label{Fig::QST_rydberg} 
\end{figure}

\subsubsection*{Code walk-through}
We perform QST on the Rydberg atoms data using the Python package {\it NetKet}~\cite{netket}. First, we import the library and define the relevant parameters for the numerical experiments. For instance, we consider the reconstruction of a square array with linear size $L=8$ (containing $N=64$ spins), with Hamiltonian parameters $V=3.0$, $\Omega=1.0$ and $\delta=2.0$.

\begin{lstlisting}[language=Python,numbers=none]
from mpi4py import MPI
import netket as nk

# Rydberg Hamiltonian parameters
L = 8        # linear size
V = 3.0      # Van der Waals interaction
Omega = 1.0  # Rabi frequency
delta = 2.0  # detuning
\end{lstlisting}

We then define the lattice structure and the Hilbert space for the neural-network wavefunction, and load the training data from file:
\begin{lstlisting}[language=Python,numbers=none]
# define the lattice structure
square_lattice = nk.graph.Hypercube(
    length=L, 
    n_dim=2, 
    pbc=False)

# build the Hilbert space
hilbert = nk.hilbert.Qubit(graph=square_lattice)

# load the training data
data_path = ".../path_to_data"
rotations, samples, bases = LoadData(hilbert, data_path)
\end{lstlisting}
Note that, since we are training in a single measurement basis, the arrays \texttt{bases}  and \texttt{rotations} are ``trivial''. Otherwise, these variables would contain respectively an integer encoding of each distinct basis, and its associated (local) unitary rotations. 

Next, we define the main components of the QST algorithm: the neural-network wavefunction, the sampler used to approximate the gradients, and the optimizer for the parameter updates. 
\begin{lstlisting}[language=Python,numbers=none]
# Neural-network wavefunction
rbm = nk.machine.RbmSpinReal(hilbert=hilbert, alpha=1)

# Monte Carlo sampler
sa = nk.sampler.MetropolisLocal(machine=rbm)

# Optimizer
op = nk.optimizer.AdaDelta(rho=0.95, epscut = 1.0e-7)

# Initialize tomography object
qst = nk.Qsr(
        sampler = sa,
        optimizer = op,
        n_samples_data = 1000,
        n_samples = 2000,
        rotations = rotations,
        samples = samples,
        bases = bases,
        sr = None)
        
\end{lstlisting}
In the definition of the RBM (with real-valued network parameters), the parameter $\alpha = n_h/N$ represents the density of hidden units. We use the AdaDelta optimizer~\cite{2012arXiv1212.5701Z} and a Metropolis sampler using simple single-spin flips. The tomography parameters \texttt{n\_samples\_data} and \texttt{n\_samples} refer respectively to the number of training samples and the number of samples drawn from the model distribution to compute the gradients for a single parameter update (i.e. for batch gradient descent). We refer the reader to the NetKet documentation for additional details. Finally, we generate the Rydberg Hamiltonian to be measured during the learning, and run the QST for a fixed number of training iterations (\texttt{epochs}).
\begin{lstlisting}[language=Python,numbers=none]
# define observable for measurements
H = generatehamiltonian(hilbert, L, L, V, Omega, delta)
qst.add_observable(H, "H")

# run quantum state tomography
for ep in qst.iter(epochs):
    obs = qst.get_observable_stats()
\end{lstlisting}

We perform QST on datasets of projective measurements generated using the MPS obtained from DMRG at different detunings. Each data set contains $10^5$ measurements. We train each RBM separately using the hyper-parameters reported above, and measure at each training iteration various observables of interest. We show the results in Fig.~\ref{Fig::QST_rydberg}, where we plot the average energy per spin $\langle\hat{H}\rangle/N$ and the average magnetizations $\langle\hat{S}^{z/x}\rangle = \sum_j\langle\hat{S}_j^{z/x}\rangle/N$ after the training has converged. Each data point is obtained by averaging the expectation values of the observables over the last 100 iterations of the training. The reconstruction show an overall good agreement with the exact values computed using the MPS wavefunctions.

\begin{figure}[t]
\noindent \centering{}\includegraphics[width=\columnwidth]{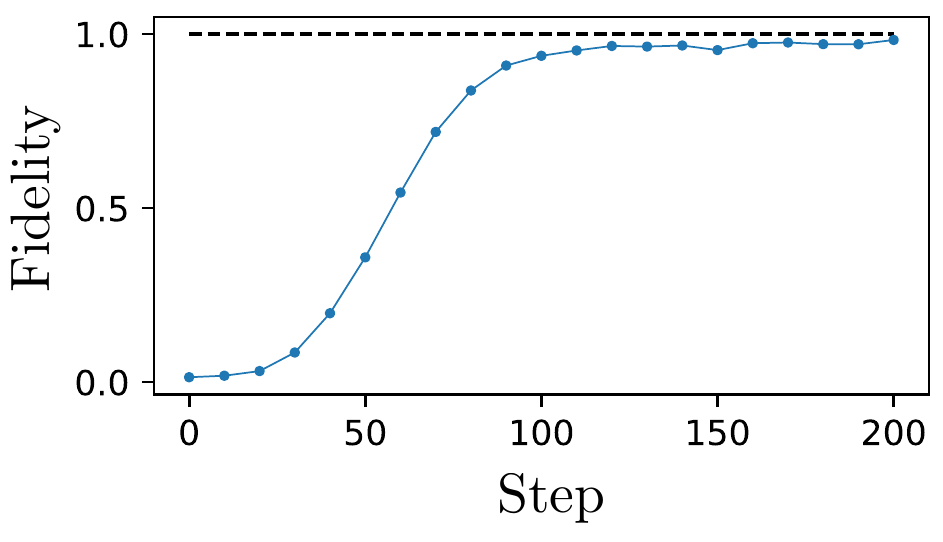}
\caption{Quantum state tomography of the Beryllium Hydride molecule. We show the fidelity $\mathcal{F} = |\langle\Phi|\psi_{\bm\theta}\rangle|^2$ between the neural-network wavefunction $\psi_{\bm\theta}$ and the exact molecular wavefunction $\Phi$ at each iteration during training.}
\label{Fig::QST_beryllium} 
\end{figure}

\subsection{Reconstruction of a molecular wavefunction}
We now move to the more general case of a wavefunction that is non-positive or complex-valued. In order to accommodate this different setup, we first need to modify the variational ansatz to allow for complex-valued amplitudes $\psi_{\bm\theta}(\bm\sigma)$. One way to achieve this is to use the RBM to parametrize the probability distribution in the reference basis (as before), and couple it with an additional RBM that parametrizes the phases, $\psi_{\bm\theta\bm\mu}(\bm\sigma)=\sqrt{p_{\bm\theta}(\bm\sigma)}e^{i\log p_{\bm\mu}(\bm\sigma)}$~\cite{torlai_Tomo}. A different strategy, which we adopt here, is to promote the weights and biases to have complex values~\cite{Carleo_2017}. Note that, for the latter representation, we cannot interpret the RBM as a probabilistic graphical model anymore. 

The presence of phases in the target wavefunction leads to an additional overhead in the measurement requirements. In fact, projective measurements in the computational basis do not carry enough information to uniquely identify the state. In order to obtain information about the phases, measurement in additional bases are required. To account for this, we can write the training data set as $\mathcal{D}=\{\bm{x}_k\}$, where each single-shot measurement is given by $\bm{x}=(\bm{\tau},\bm{\sigma})$, with $\bm\tau$ and $\bm\sigma$ referring respectively to the measurement basis and the binary measurement outcome. For example, for the case of Pauli measurements, the data point $\bm{x} = (\sigma_1^x=0,\sigma^z_2=1)$ refers to a measurement of qubit 1 (2) in the eigenbasis of the Pauli-X (Z) operator.

The learning algorithm proceeds in a similar fashion to the case of a positive wavefunction, reducing to the optimization of the negative log-likelihood cost function
\begin{equation}
\mathcal{C}(\bm{\theta})=-\frac{1}{|\mathcal{D}|}\sum_{k=1}^{|\mathcal{D}|}\log p_{\bm{\theta}}(\bm{x}_k)\:.
\end{equation}
The important difference is that, in order to evaluate measurement probabilities in bases other than the reference one, we need to appropriately rotate the neural-network wavefunction. We can write this measurement probability as
\begin{equation}
p_{\bm{\theta}}(\bm{x}_k) = \frac{|\langle\bm{\sigma}_k|\bm{U}(\bm{\tau}_k)|\psi_{\bm\theta}\rangle|^2}{\langle\psi_{\bm\theta}|\psi_{\bm\theta}\rangle}\:,
\end{equation}
where $\bm{U}(\bm{\tau})$ refers to the unitary transformation that rotates the reference basis into the measurement basis $\bm{\tau}$. We also assume local measurements, leading to the factorization $\bm{U}(\bm{\tau})=\bigotimes_{j=1}^NU(\tau_j)$ into single-qubit unitary rotations $U(\tau_j)$. For the example described above, with $\linebreak\bm{x} = (\sigma_1^x=0,\sigma^z_2=1)$, the neural-network probability is (up to a normalization factor) $p_{\bm\theta}(\bm{x})\propto|\langle01|H_1\otimes\mathbb{1}_2|\psi_{\bm\theta}\rangle|^2$, where $H$ is the Hadamard gate (i.e. the rotation into the $\sigma^x$ basis).

We show neural-network QST of a non-positive wavefunction for the case of the electronic ground state of a small molecule. The Hamiltonian in second quantization is given by
\begin{equation}
\hat{H} = \sum_{\alpha,\beta}t_{\alpha\beta}\hat{c}^\dagger_\alpha \hat{c}_\beta + \frac{1}{2}\sum_{\alpha,\beta,\gamma,\delta}u_{\alpha\beta\gamma\delta} \hat{c}^\dagger_\alpha \hat{c}^\dagger_\beta \hat{c}_\gamma \hat{c}_\delta
\end{equation}
where $\hat{c}^\dagger$ and $\hat{c}$ are fermionic creation and annihilation operators, and $t_{\alpha\beta}$ and $u_{\alpha\beta\gamma\delta}$ are electronic integrals. This Hamiltonian can be mapped into a qubit Hamiltonian using one of several mappings (i.e. Jordan-Wigner, Bravyi-Kitaev etc)
\begin{equation}
\hat{H} = \sum_k c_k \hat{P}_k\:.
\end{equation}
where $c_k$ are interaction coefficients and $\hat{P}_k$ are operators that belongs to the $N$-qubit Pauli group. 

We specifically look at the Beryllium Hydride molecule (\BeH) in the STO-3G basis, mapped to $N=6$ qubits~\cite{kandala_hardware-efficient_2017}. In order to generate the training data, we first obtain the full ground state wavefunction $|\Phi\rangle$ with exact diagonalization of the qubit Hamiltonian. Given the ground state, we can generate measurement data by sampling the full probability distribution obtained from the Born rule. A single measurement is obtained by first selecting a measurement basis $\bm\tau$, then rotating the wavefunction accordingly $|\Phi_{\bm\tau}\rangle=\bm{U}(\bm\tau)|\Phi\rangle$, and finally sampling the measurement outcome $\bm\sigma\sim P_{\bm\tau}(\bm{\sigma})=|\langle\bm\sigma|\Psi_{\bm\tau}\rangle|^2$. We choose the measurement bases according to the Pauli operators $\hat{P}_k$ appearing the Hamiltonian, each one being selected randomly among this set. We show the results of the QST experiment in Fig.~\ref{Fig::QST_beryllium}, where we plot the fidelity between the neural-network wavefunction and the target wavefunction during the training.

\subsection{Discussion}
We have shown that for wavefunctions whose amplitudes can be gauged to be real and positive, QST is equivalent to unsupervised learning of projective measurement data in a single basis. For more general wavefunctions with a sign structure or complex-valued amplitudes, measurements in multiple bases are required to reconstruct the phases. These are processed by the neural network by appropriately rotating the parametrized wavefunction with a unitary $\bm{U}(\bm{\tau})=\bigotimes_{j=1}^NU(\tau_j)$ composed by single-qubit basis rotations $U(\tau_j)$. In practice, this operation entails an exponential cost in the number of non-trivial rotations $U(\tau_j)\ne\mathbb{1}_j$, which means that only a small fraction of any IC set of bases can be use to train the neural network. Nevertheless, this reduced amount of information may be enough to reconstruct sufficiently structured quantum states. For example, a ground state of a local gapped Hamiltonian can be identified by the statistics of projective measurements in a set of bases corresponding to the decomposition of the Hamiltonian in the Pauli group.

An important assumption that was made is the purity of the quantum state under reconstruction, which is violated in any practical setting. There are cases where an approximate pure state reconstruction of an experimental quantum state  may be justified, and could still provide valuable insights~\cite{torlai_rydberg19}. However, for benchmarking and noise characterization tasks, one needs to reconstruct the full density matrix. This can be achieved by introducing a neural-network parametrization of a density operator $\rho_{\bm\theta}(\bm\sigma,\bm\sigma^\prime)$, which is trained in an analogous way. The positivity of $\rho_{\bm\theta}$ can be enforced using a purification scheme, where $\rho_{\bm\theta}$ is purified by additional latent units in the neural network~\cite{Torlai_latent}. An iterative procedure to discover the dominant eigenstates of a density operators using RBMs has also been put forward~\cite{PhysRevA.102.022412}.

A different approach for reconstructing generic mixed quantum states consists of directly parametrizing the probability distribution $p(\bm\alpha) = \tr[\bm{\varrho}\:\Pi_{\bm\alpha}]$ of an IC-POVM set $\{\Pi_{\bm\alpha}\}$~\cite{carrasquilla_povm}. Contrary to the purification scheme, this approach does not require unitary rotations, and thus lifts the exponential complexity associated with processing data from arbitrary local bases. This however comes at the cost of possibly violating the positivity of the learned density matrix, since this cannot be enforced at the level of the POVM distribution parametrization.

%----------------------------------------------------------------------------------------
%----------------------------------------------------------------------------------------
%                               VARIATIONAL MONTE CARLO
%----------------------------------------------------------------------------------------
%----------------------------------------------------------------------------------------
%----------------------------------------------------------------------------------------

\section{Variational ground state optimization}
\label{vmc}
The variational principle in quantum mechanics states that the expectation value of the Hamiltonian of a physical system of interest over any valid wavefunction is always greater than or equal to the ground state energy of the system. This principle indicates that a strategy for finding an approximation to the ground state energy of a system is to start from a parameterized wavefunction and vary its parameters until it yields the minimum possible energy. 

Historically, the choice of wavefunction, which is critical to the success of the algorithm, has traditionally been made motivated in close connection to a physical understanding of the problem, e.g., from approximate mean-field solutions supplemented with some form of additional correlation such as a Jastrow factor~\cite{becca_sorella_2017}. More recently, motivated by their representation power, efficiency, and generality, neural networks have been explored as trial wavefunctions~\cite{Carleo_2017}.  In particular, and as we explore below, recurrent neural networks are naturally well-suited to the study of systems exhibiting strong correlations such as those arising in the study of classical and quantum systems, which are prevalent in condensed matter and statistical physics~\cite{Wu_2019,carrasquilla_povm,RNNWF_2020, roth2020iterative}.

\subsection{Recurrent neural networks}
\begin{figure}[t]
\noindent \centering{}\includegraphics[width=\columnwidth]{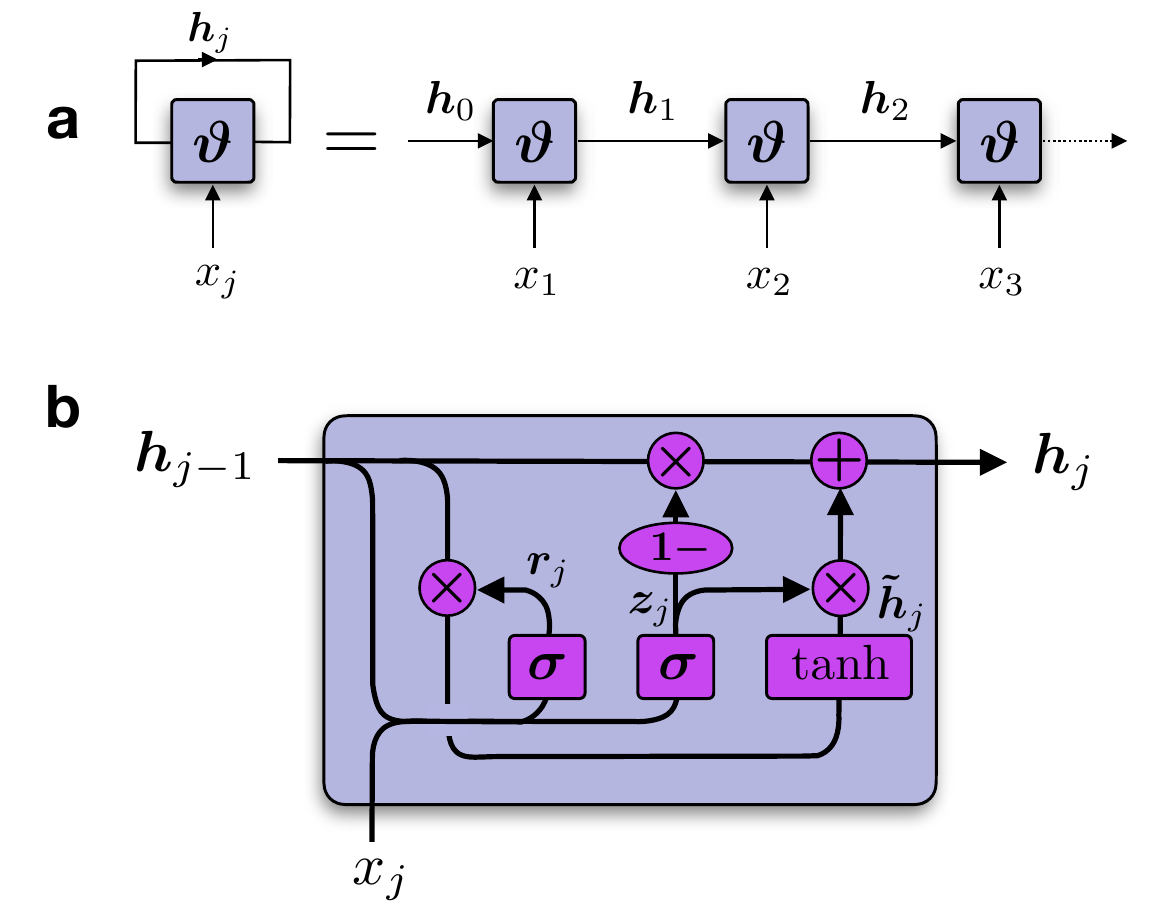}
\caption{Recurrent neural network (RNN). {\bf a}). A generic RNN cell (left), and its unrolling in time to process sequenced data (right). {\bf b}) The gated recurrent unit (GRU), and the set of operations used to process the input latent state $\bm{h}_{j-1}$ and visible state $x_j$ to generate a new latent state $\bm{h}_j$. Lines joining together means concatenation, while the circles are element-wise operations.}
\label{Fig::rnn} 
\end{figure}

A recurrent neural network (RNN) models a probability distribution $p(\bm{\sigma})=p(\sigma_1,\dots,\sigma_N)$ using a sequential structure according to the chain rule of probabilities
\begin{equation}
p(\bm{\sigma})=p(\sigma_1)\:p(\sigma_2\,|\,\sigma_1)\:\prod_{j=3}^Np(\sigma_j\,|\bm{\sigma}_{-j}),
\end{equation}
where $\bm{\sigma}_{-j}=(\sigma_1,,\dots,\sigma_{j-2},\sigma_{j-1})$. The recurrent unit in the neural network parametrizes the conditional probability distribution $p(\sigma_j\,|\,\bm{\sigma}_{-j})$ at any {\it time-step} $j$, and it processes the data according to the input $\sigma_{j-1}$ and a recurrent latent vector $\bm{h}_j$ (Fig.~\ref{Fig::rnn}a). 

We specifically consider Gated Recurrent Units (GRU)~\cite{cho-etal-2014-learning}, introduced to solve the vanishing gradient issue of vanilla RNNs. The schematic of a GRU units is shown in (Fig.~\ref{Fig::rnn}b). Given a {\it visible} input state $\sigma_{j-1}$ and a latent state $\bm{h}_{j-1}$, the GRU unit at time-step $j$ processes them according to the sequence of operations and outputs an updated latent vector $\bm{h}_j$. The first two operations are the so-called {\it update gate} and {\it reset gate}:
\begin{align}
\bm{z}_j = \text{sig}\Big(\bm{W}_z\:[\bm{h}_{j-1}; \sigma_{j-1}]+\bm{b}_z\Big)\\
\bm{r}_j = \text{sig}\Big(\bm{W}_r\:[\bm{h}_{j-1}; \sigma_{j-1}]+\bm{b}_r\Big)
\end{align}
where $[\bm{h}_{j-1}; \sigma_{j-1}]$ is the concatenation operation, $\text{sig}(x)=(1+e^{-x})^{-1}$ is the sigmoid function, and $\bm{W}_z$, $\bm{W}_r$ and $\bm{b}_z$, $\bm{b}_r$ are respectively weights and biases variational parameters. These gates are used to control how much information about previous time-steps is kept encoded into the latent vector. 

Next, given the input visible state $x_j$ at the current time-step, an internal latent state is created according to
\begin{equation}
\bm{\tilde{h}}_j = \tanh\Big(\bm{\tilde{W}}\:[\bm{r}_j\odot\bm{h}_{j-1},x_j]+\bm{\tilde{b}}\Big)
\end{equation}
where $\bm{a}\odot\bm{b}$ is the element-wise multiplication, and $\bm{\tilde{W
}}$, $\bm{\tilde{b}}$ are a new set of parameters. The new latent vector (output of the GRU), is generated as
\begin{equation}
\bm{h}_j=(\bm{1}-\bm{z}_j)\odot\bm{h}_{j-1}+\bm{z}_j\odot\bm{\tilde{h}}_j\:,
\end{equation}
and sent to the GRU unit at time-step $(j+1)$. Finally, the conditionals are computed using a softmax layer
\begin{equation}
    p(\sigma_j \,|\,\bm{\sigma}_{-j}) = S\left( U \bm{h}_j + \bm{c}\right)
\end{equation}
where $U$ and $\bm{c}$ are the parameters of the softmax layer. 

Given the above parametrization of a probability distributions $p(\bm{\sigma})$, we can now promote RNNs to quantum mechanical wavefunctions $\psi(\bm{\sigma})$. As noted in the QST examples, we stress that {\it stoquastic} many-body
Hamiltonians have ground states $\ket{\Psi}$ with strictly real and positive amplitudes in the standard computational basis~\cite{Bravyi2006}.
This class states can be represented in terms of probability distributions,
\begin{align}
    \ket{\Psi} = \sum_{\bm{\sigma}} \psi(\bm{\sigma})\ket{\bm{\sigma}} = \sum_{\bm{\sigma}}\sqrt{p(\bm{\sigma})}\ket{\bm{\sigma}}.
\end{align}
% We note that this property has been exploited in multiple wavefunction representations using generative models such as restricted Boltzmann machines~\cite{Carleo_2017}.
For such family of quantum states, which includes the Rydberg system considered in this work, it is natural to try to approximate $p(\bm{\sigma})$ with an RNN.

\subsection{Variational Monte Carlo simulation of Rydberg atoms}
% \textcolor{red}{JC: Let's write the VMC part here, and just show the results at the end.}
The goal of variational Monte Carlo (VMC) is to iteratively optimize an ansatz wavefunction to approximate  ground states of local Hamiltonians~\cite{becca_sorella_2017}. The VMC method make use of trial wavefunction $\ket{\Psi_{\bm{\theta}}}$ endowed with parameters $\bm{\theta}$. Here we consider an GRU-RNN wavefunction. Crucially, we exploit the fact that the RNN wavefunction allows for efficient sampling from the square of the amplitudes of $\ket{\Psi_{\bm{\theta}}}$. 

The  VMC method iteratively optimizes the expectation value of the
energy
\begin{align}
  E \equiv \frac{\braket{\Psi_{\bm{\theta}}|\hat{H}|\Psi_{\bm{\theta}}}}{\braket{\Psi_{\bm{\theta}}|\Psi_{\bm{\theta}}}}.\label{eq:energy}
\end{align}
The minimization is carried out using the gradient descent method or any other variant of it. Since the RNN wavefunction is normalized such that $\braket{\Psi_{\bm{\theta}}|\Psi_{\bm{\theta}}}=1$, the expectation value in Eq.~\ref{eq:energy} can be written as
\begin{align}
  E = \braket{\Psi_{\bm{\theta}} |\hat{H}|\Psi_{\bm{\theta}}}
  &=\sum_{\bm{\sigma}} |\psi_{\bm{\theta}}(\bm{\sigma})|^2\sum_{\bm{\sigma'}} H_{\bm{\sigma\sigma'}}\frac{\psi_{\bm{\theta}}(\bm{\sigma'})}{\psi_{\bm{\theta}}(\bm{\sigma})} \nonumber \\
  &\equiv \sum_{\bm{\sigma}} |\psi_{\bm{\theta}}(\bm{\sigma})|^2E_{loc}(\bm{\sigma}) \nonumber \\
   &\approx \frac{1}{N_S}\sum_{\bm{\sigma} \sim |\psi_{\bm{\theta}}(\bm{\sigma})|^2} E_{loc}(\bm{\sigma}),
  \label{eq:expectation_value} 
\end{align}
which represents a sample average of the local energy $E_{loc}(\bm{\sigma})$. The gradients $\partial_{{\bm{\theta}}}E$ can be similarly written as
\begin{align}
  \partial_{{\bm{\theta}}} E =
  \sum_{\bm{\sigma}} |\psi_{\bm{\theta}}(\bm{\sigma})|^2\frac{\partial_{{\bm{\theta}}}\psi^{*}_{\bm{\theta}}(\bm{\sigma})}{\psi^{*}_{\bm{\theta}}(\bm{\sigma})} E_{loc}(\bm{\sigma})  + \text{c.c}.
  \label{eq:gradient}
\end{align}
An optimization step involves drawing $N_S$ samples $\{ \bm{\sigma}^{(1)}, \bm{\sigma}^{(2)}, \ldots, \bm{\sigma}^{(N_S)}\}$ from $|\psi_{\bm{\theta}}(\bm{\sigma})|^2$, followed by an estimation of the energy gradients 
\begin{align}
\partial_{{\bm{\theta}}} E \approx
\frac{2}{N_S} \mathfrak{Re} \left ( \sum_{i=1}^{N_S}  \frac{\partial_{{\bm{\theta}}}\psi^{*}_{\bm{\theta}}(\bm{\sigma^{(i)}})}{\psi^{*}_{\bm{\theta}}(\bm{\sigma^{(i)}})} E_{loc}(\bm{\sigma^{(i)}}) \right),
\label{eq:stochastic_gradient1}
\end{align}
using automatic differentiation~\cite{AD_2017} and updating the parameters according to
\begin{align}
  {\bm{\theta}} \leftarrow {\bm{\theta}} - \alpha \partial_{{\bm{\theta}}}E
  \label{eq:gradient_descent}
\end{align}
with a small learning rate $\alpha$. Instead of this simple gradient descent rule, we can also use the Adam optimizer \cite{2014arXiv1412.6980K} to implement the parameter updates.

We note that the stochastic evaluation of the gradients in Eq.~\eqref{eq:stochastic_gradient1} implies that these may exhibit high variance, which can potentially slow down the convergence of the algorithm. This problem can be alleviated through the introduction of a term in Eq.~\eqref{eq:stochastic_gradient1} that helps reduce the variance of the gradients~\cite{RNNWF_2020}
\begin{align}
\partial_{\bm{\theta}} E &\approx
\frac{2}{N_S} \mathfrak{Re} \left ( \sum_{i=1}^{N_S}  \frac{\partial_{\bm{\theta}}\psi^{*}_{\bm{\theta}}(\bm{\sigma^{(i)}})}{\psi^{*}_{\bm{\theta}}(\bm{\sigma^{(i)}})} \left ( E_{loc}(\bm{\sigma^{(i)}}) - E \right ) \right) \nonumber\\
 &=
\frac{2}{N_S} \mathfrak{Re} \left ( \sum_{i=1}^{N_S} \partial_{{\bm{\theta}}} \log \psi^{*}_{\bm{\theta}}(\bm{\sigma^{(i)}}) \left ( E_{loc}(\bm{\sigma^{(i)}}) - E \right ) \right).
\label{eq:stochastic_gradient2}
\end{align}
This estimator has improved variance, stabilizes the convergence of the algorithm, and is unbiased~\cite{RNNWF_2020}. We note that in the limit where $E_{loc}(\bm{\sigma^{(i)}}) \approx E$ near convergence, the variance of the gradients $\partial_{{\bm{\theta}}} E$ goes to zero as opposed to the nonzero variance of the gradients in Eq.~\eqref{eq:stochastic_gradient1}.

\subsubsection*{Code walk-through}
We provide an implementation of VMC using a RNN wavefunction based on the TensorFlow library. We first define all the relevant parameters of the Rydberg Hamiltonian and the RNN training.
\begin{lstlisting}[language=Python,numbers=none]
# Hamiltonian parameters
Lx = 4      # linear size in x direction
Ly = 4      # linear size in y direction
N = Lx*Ly   # total number of atoms
V = 7.0     # Van der Waals interaction
Omega = 1.0 # Rabi frequency
delta = 1.0 # detuning 

# RNN-VMC parameters
lr = 0.001     # learning rate
nh = 32        # number of hidden units
ns = 500       # number of samples
epochs = 1000  # training iterations
seed = 1234    # seed of RNG

# initialize the RNN wavefunction
vmc = VariationalMonteCarlo(
         Lx, Ly, V, Omega, delta,
         nh, lr, epochs, seed)
\end{lstlisting}
The RNN parameters are the learning rate for the optimizer, the number of hidden units in the GRU cell, the number of samples used to approximate the energy (and its gradients) at each training iteration, and the total number of epochs. The VMC module is then initialized accordingly.

We can now perform the VMC simulation by training the RNN parameters. For a given number of epochs, we first sample the RNN distribution to generate \texttt{ns} samples. These can be used to evaluate the cost function of the optimization and its gradients, computed here using the AD functionalities from TensorFlow. The parameters are then updated according to the gradients.
\begin{lstlisting}[language=Python,numbers=none]
# training loop
for n in range(epochs):
    # sample the RNN wavefunction
    samples, _ = vmc.sample(ns)
    
    # evaluate the loss function in AD mode
    with tf.GradientTape() as tape:
        logpsi = vmc.logpsi(samples)
        eloc = vmc.localenergy(samples, logpsi)
        Eo = tf.stop_gradient(tf.reduce_mean(eloc))

        loss = tf.reduce_mean(2.0*tf.multiply(logpsi, tf.stop_gradient(eloc)) - 2.0*Eo*logpsi)
    
    # compute the gradients
    gradients = tape.gradient(loss, vmc.trainable_variables)
    
    # update the parameters
    vmc.optimizer.apply_gradients(zip(gradients, vmc.trainable_variables))
    
    # get average energy	
    avg_E = np.mean(eloc.numpy())
\end{lstlisting}

\begin{figure}[t]
\noindent \centering{}\includegraphics[width=\columnwidth]{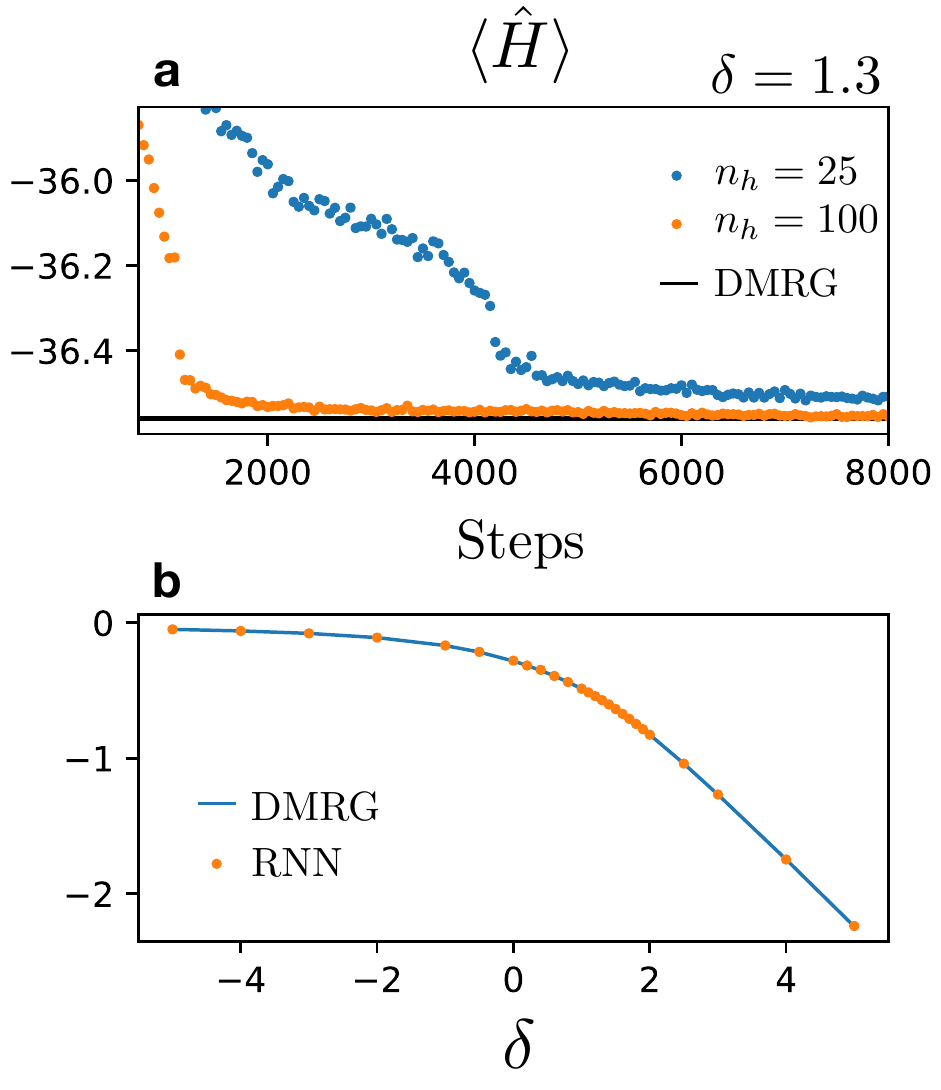}
\caption{Variational Monte Carlo simulation of the Rydberg Hamiltonian. ({\bf a}) Average energy during as a function of the training step near the phase transition at $\delta=1.3$. We show RNN wavefunctions with $n_h=25$ and $n_h=100$ hidden units, compared to the result obtained from DMRG. ({\bf b}) Average energy over the full phase diagram for a RNN wavefunction with $n_h=100$, in comparison with DMRG.}
\label{Fig::vmc} 
\end{figure}

We show in Fig.~\ref{Fig::vmc} the results of VMC simulations for a $8\times8$ array of Rydberg atoms. In Fig.~\ref{Fig::vmc}(a), we plot the average total energy at each training iteration for a RNN wavefunction with a GRU containing $n_h=25$ and $n_h=100$ hidden units. As expected, increasing the number of hidden units in the RNN leads to a improved accuracy in the training. In Fig.~\ref{Fig::vmc}(b) we plot the average energy after convergence as a function of the detuning using $n_h=100$, and compare with the values obtained from DMRG.

\section{Conclusions}
In this Article, we presented applications of machine learning algorithms based on neural networks to quantum many-body physics. We focused on numerical demonstrations and provided hands-on code tutorials based on open-source software~\cite{tensorflow,netket,itensor}, with the goal of facilitating learning for researcher new to the field, and accelerating the adoption of machine learning in  quantum physics.

We showcased distinct machine learning paradigms, implemented with different neural-network architectures. As a testbed for the numerical experiments, we have chosen a system of interacting Rydberg atoms arranged in a two-dimensional square array. First, we demonstrated supervised learning of atomic occupation data, which was generated from ground states of the Rydberg Hamiltonian obtained using the density-matrix renormalization group. Using a convolutional neural network trained on labelled data, we showed how to learn the quantum phase transition between a disordered phase and the anti-ferromagnetically ordered phase. 

We presented unsupervised learning of unlabelled data using the restricted Boltzmann machine. We showed that for pure quantum states with real and positive amplitude (e.g. the Rydberg ground states), this procedure is equivalent to quantum state tomography based on a neural-network representation of a quantum state. Using atomic occupation data, we trained Boltzmann machines to learn the ground states of the Rydberg Hamiltonian. We also described a simple extension of this approach to learn quantum states with a sign structure, and showed a demonstration in the context of simulation of chemistry with quantum computers, where we learned the ground state of a molecule using qubit measurements.

The final application we explored is the Monte Carlo optimization of a variational wavefunction to estimate the ground state of a many-body Hamiltonian. We parametrized a wavefunction using a recurrent neural-network and trained its parameters to lower both the expectation value and the variance of the Rydberg Hamiltonian. 

It is becoming increasingly evident that machine learning is full of thrilling opportunities and conceptual advances with great potential to energize computational and experimental physics. As these notions continue to spread through the research landscape of strongly-correlated quantum matter and quantum information science, we hope this tutorial will provide a useful first step into the expanding domain of artificial intelligence for the study of quantum many-body systems.
   
\section*{Acknowledgements}
The numerical simulation were performed on the Simons Foundation Super-Computing Center, using the following software libraries: ITensor~\cite{itensor} (for the simulation of the Rydberg atoms), TensorFlow~\cite{tensorflow} (for supervised learning and variational Monte Carlo), and NetKet~\cite{netket} (for quantum tomography). The Flatiron Institute is supported by the Simons Foundation. JC acknowledges support from Natural Sciences and Engineering Research Council of Canada (NSERC), the Shared Hierarchical Academic Research Computing Network (SHARCNET), Compute Canada, Google Quantum Research Award, and the Canadian Institute for Advanced Research (CIFAR) AI chair program. 

\bibliography{bibliography}

\end{document}